\documentclass[usenatbib]{mnras}
\usepackage{graphicx}
\usepackage{amsmath}
\usepackage{wrapfig}
\usepackage{lscape}
\usepackage{array}
\usepackage{tabularx}
\usepackage{tablefootnote}
\usepackage{soul} %Strikethrough
\usepackage[dvipsnames]{xcolor} %highlight
\usepackage[flushleft]{threeparttable}
 
\usepackage{subfigure}
\usepackage{xcolor}
\usepackage{color, soul}
\graphicspath{ {images/} }

\begin{document}
%%%%%%%%%%%%%
%%% TITLE %%%
%%%%%%%%%%%%%
\title[Constraining the CSFH from the improved EBL]{GAMA/DEVILS: Constraining the cosmic star-formation history from improved measurements of the 0.3-2.2$\mu$m  Extragalactic Background Light}
%evolution, from improved Cosmic Optical Background measurements of the EBL}

\author[S. Koushan et. al.]{Soheil Koushan$^{1}$\thanks{E-mail:
 so.koushan@gmail.com}, Simon P. Driver$^{1}$, Sabine Bellstedt$^{1}$,  Luke J.~Davies$^{1}$,  \newauthor Aaron S. G. Robotham$^{1}$, Claudia del P Lagos$^{1,2}$, Abdolhosein Hashemizadeh$^{1}$, 
 \newauthor Danail Obreschkow$^{1}$, Jessica E. Thorne$^{1}$, Malcolm Bremer$^{3}$, B.W. Holwerda$^{4}$, \newauthor Andrew M. Hopkins$^{5}$, Matt J. Jarvis$^{6}$, Malgorzata Siudek$^{7,8}$, and 
 \newauthor Rogier A. Windhorst$^{9}$\\
$^1$ICRAR, The University of Western Australia, 35 Stirling Highway, Crawley WA 6009, Australia\\
$^2$ARC Centre of Excellence for All Sky Astrophysics in 3 Dimensions (ASTRO 3D)\\
$^3$Astrophysics Group, School of Physics, University of Bristol, Tyndall Avenue, Bristol, BS8 1TL, UK\\
$^4$Department of Physics and Astronomy, 102 Natural Science Building, University of Louisville, Louisville KY 40292, USA\\
$^5$Australian Astronomical Optics, Macquarie University, 105 Delhi Rd, North Ryde, NSW 2113, Australia\\
$^{6}$ Oxford Astrophysics, Denys Wilkinson Building, Keble Road, Oxford OX1 3RH, UK\\
$^{7}$ Institut de Fisica d'Altes Energies (IFAE), The Barcelona Institute of Science and Technology, 08193 Bellaterra (Barcelona), Spain\\
$^{8}$ National Centre for Nuclear Research, ul. Hoza 69, 00-681 Warsaw, Poland\\ 
$^{9}$School of Earth \& Space Exploration, Arizona State University, Tempe, AZ85287-1404, USA\\
}
%$^\dagger${E-mail: soheil.koushan@icrar.org} 

% These dates will be filled out by the publisher
\date{Accepted XXX. Received YYY; in original form ZZZ}

% Enter the current year, for the copyright statements etc.
\pubyear{2021}

\label{firstpage}
\pagerange{\pageref{firstpage}--\pageref{lastpage}}
\maketitle
%%%%%%%%%%%%%%%%
%%% ABSTRACT %%%
%%%%%%%%%%%%%%%%
\begin{abstract}
We present a revised measurement of the optical extragalactic background light (EBL), based on the contribution of resolved galaxies to the integrated galaxy light (IGL). The cosmic optical background radiation (COB), encodes the light generated by star-formation, and provides a wealth of information about the cosmic star formation history (CSFH). We combine wide and deep galaxy number counts from the Galaxy And Mass Assembly survey (GAMA) and Deep Extragalactic VIsible Legacy Survey (DEVILS), along with the Hubble Space Telescope (HST) archive and other deep survey datasets, in 9 multi-wavelength filters to measure the COB in the range from 0.35 $\mu$m to 2.2 $\mu$m. We derive the luminosity density in each band independently and show good agreement with recent and complementary estimates of the optical-EBL from very high-energy (VHE) experiments. Our error analysis suggests that the IGL and $\gamma$-ray measurements are now fully consistent to within $\sim$10$\%$, suggesting little need for any additional source of diffuse light beyond the known galaxy population. We use our revised IGL measurements to constrain the cosmic star-formation history, and place amplitude constraints on a number of recent estimates. 
As a consistency check, we can now demonstrate convincingly, that the CSFH, stellar mass growth, and the optical-EBL provide a fully consistent picture of galaxy evolution. We conclude that the peak of star-formation rate lies in the range $0.066-0.076$ M$_{\bigodot}$yr$^{-1}$ Mpc$^{-3}$ at a lookback time of 9.1 to 10.9 Gyrs.

%We also explore the prospect of constraining the mean gas-phase metallicity evolution through the shape of the optical-EBL. 
%and that the current day mean gas-phase metallicity is between 0.016 and 0.028. 
\end{abstract}
% We provide an electronic version of our best fit COB/CIB EBL model for use in VHE analyses.
% , and show that while constraints are currently weak this approach provides a promising pathway forward as our measurements improve
%%%%%%%%%%%%%%%%
%%% KEYWORDS %%%
%%%%%%%%%%%%%%%%
% methods: data analysis
% galaxies: evolution
% galaxies: statistics
% cosmology: cosmic background radiation
% cosmology: observations

%%%%%%%%%%%%%%%%%%%%
%%% INTRODUCTION %%%
%%%%%%%%%%%%%%%%%%%%
\section{Introduction}
The extragalactic background light (EBL) is defined as the total flux received today (nWm$^{-2}{\rm steradian}^{-1}$) from all sources of photon production since the epoch of recombination plus the Cosmic Microwave Background (CMB). The CMB component is somewhat distinct, as it represents the radiation field from the epoch of recombination as the temperature, pressure, and density of the Universe dropped to the point where nuclei and electrons combined to form neutral atoms (\citealp{2001ARA&A..39..249H}; \citealp{2006NewAR..50..208K}; \citealp{2009astro2010S..54C}; \citealp{2009A&A...505.1041G}; \citealp{2011MNRAS.410.2556D}; \citealp{2013ApJ...768..197I}; \citealp{2016ApJ...827..108D}). Hence, the EBL represents the integrated luminosity cone of the universe from the CMB surface to the present day. Understanding this energy is essential as it encodes all photon production pathways. This photon-energy production pervades the entire electromagnetic spectrum, but is typically broken into distinct wavelength `windows' (see \citealp{2018ApSpe..72..663H}), aligned with the detection technologies. \\

Overall, the CMB, the redshifted thermal remnant of the hot early universe \citep{1965ApJ...142..419P}, is the dominant component of the EBL. The subsequent photon production represents the remainder of the EBL, and has been produced by stars, gas, accreting black holes (BHs) and dust emission --- and other processes --- arising over the entire path length of the Universe (i.e. since $\sim$ 380,000 years after the big bang). The CMB also dominates the energy distribution of the EBL, being a factor of $\sim$40 times higher ($\sim$960 nWm$^{-2}sr^{-1}$) than either the infrared or optical backgrounds (\citealp{2000ASPC..201..403S}; \citealp{2004NewAR..48..465W}; \citealp{2006A&A...451..417D}; \citealp{2016ApJ...827..108D}; \citealp{2020bugm.conf....7D}).\\

The cosmic optical background (COB) light is the second most important part of the EBL in terms of photon production, and is tied with the cosmic infrared background (CIB) in terms of final photon density and integrated photon energy (NB: most CIB photon energy originates from photons produced in the COB, which are absorbed and reradiated to form the CIB, hence the COB dominates in terms of photon production). The COB is one of the least well-known components of the EBL spectra, primarily due to the uncertainties in subtracting the intervening foregrounds. Formally, the COB is defined to be the emission emitted in the wavelength range 0.1 $\mu$m to 8 $\mu$m (i.e. the UV to mid-infrared, MIR), and arises primarily from the formation and evolution of stellar mass as galaxies form and evolve (\citealp{2001ARA&A..39..249H}). Hence, the energy density of the COB is closely related to the growth of stellar mass over all time (\citealp{1999ASPC..193..487B}). There is also a non-negligible contribution from Active Galactic Nuclei (AGN) whereby light, due to the accretion onto a supermassive BH, escapes from the host galaxy and radiates through the inter-galactic medium (e.g. \citealp{2011ApJ...736..119M};  \citealp{Driver2016}).  

The COB also encodes information about the entire cosmic star formation history, as well as the origin and fate of the very first stars responsible for reionisation which constitutes 0.1 to 1$\%$ of the total EBL (\citealp{2009astro2010S..54C}; \citealp{2018ApJS..234...41W}). The reasons above make the COB one of the most highly studied portions of the background light with tremendous potential to provide strong astrophysical constraints on galaxy formation and evolution.\\

At longer wavelengths the cosmic infrared background (CIB) light is defined as emission at wavelengths between 8 $\mu$m and 1000 $\mu$m  (\citealp{1996A&A...308L...5P}; \citealp{1999A&A...344..322L}; \citealp{2001ARA&A..39..249H}; \citealp{2006NewAR..50..208K}). As mentioned, a significant amount of the starlight produced at UV and optical wavelengths,  is absorbed by dust in the host galaxies and later re-radiated in MIR to far-infrared (FIR), contributing to the CIB. The CIB spectral energy distribution (SED) has a peak of the emission at 100 $\mu$m equating to thermal dust emission with temperatures in the range 20-60K. As detected from the Earth, both the COB and CIB have approximately the same contribution of 24 and 26 nWm$^{-2}sr^{-1}$, respectively (\citealp{2016ApJ...827..108D}). Thus, the combination of the IR and optical background light equates, energy-wise, to about $\sim$5$\%$ of the CMB. \\

Unfortunately, despite significant efforts, the current measurements of the COB and CIB exhibit significant inconsistencies most likely associated with the measurement methods (see reviews by \citealp{2018ApSpe..72..663H}; \citealp{2019ConPh..60...23M}; \citealp{2020bugm.conf....7D}). Conventionally, these efforts are classified as direct (background sky brightness; \citealp{2005ApJ...626...31M}; \citealp{2012IAUS..284..482C}; \citealp{2011ApJ...736..119M}; \citealp{2013ApJS..207...31Z}; \citealp{2017NatCo...815003Z}; \citealp{2017PASJ...69...31K}), and indirect (based on counting galaxies; \citealp{2000MNRAS.312L...9M}; \citealp{2001ApJ...550L.137T}; \citealp{2005ApJ...619L..11X}; \citealp{2009A&A...505.1041G}; \citealp{2010ApJ...723...40K}; \citealp{2016ApJ...827..108D}). \\

Direct measurement is the traditional technique applied to estimate the COB, including both discrete and diffuse emission. This is based on detecting the absolute flux of the night sky from space platforms and subtracting the contributing foregrounds. The dominant foreground component outside of the Earth's atmosphere is the zodiacal dust glow. This foreground is caused by minuscule dust fragments reflecting sunlight and forming the so-called zodiacal light (ZL; \citealp{1981A&A...103..177L}; \citealp{2002ApJ...571...85B}; \citealp{2006Natur.440.1018A}). In addition to the ZL, the diffuse sky emission may contain residual airglow (unless the observing platform is entirely beyond Earth's orbit), and diffuse galactic light (DGL). The optimal technique for separating the DGL and ZL+Airglow remains controversial (\citealp{2019PASJ...71...82K}). The intensity of the DGL is 5 to 10 times larger than the EBL, and the ZL+Airglow varies from 30 to 100 times the EBL (\citealp{2012IAUS..284..429M}). These foregrounds are also time/day variable (i.e. ZL+Airglow), and so they need to be estimated for each observation. Therefore, searching for an optimised and well-established model to remove the foregrounds efficiently is of high priority. One suggested technique designed to overcome these backgrounds is the Dark Cloud method (\citealp{2017MNRAS.470.2133M}), in which the effect of the foregrounds on the background light is measured by estimating the differential measurement between a dark nebula at high galactic latitude and a surrounding clear region. This results in an upper limit to the UV and blue portion of the EBL. \\ % implies the EBL intensity

The indirect method, alternatively, represents the integrated galaxy counts from deep observations and defines the entire photon flux from discrete sources (i.e. galaxies, quasars) only. In this work, we aim to measure the integrated galaxy light from a compendium of wide and deep galaxy number count data from the optical to the near-infrared (NIR). By having sufficiently wide and deep photometric data, the EBL obtained from the galaxy count method should converge to that from direct measurements. The difference in the two methods can, in due course, provide constraints on any diffuse sources (see the discussion in \citealp{2016ApJ...827..108D}). Over the past decades, there has been a rapid increase of new data provided by e.g. \textsl{HST} (HST COSMOS; \citealp{2007ApJS..172....1S}, HST GOODs; \citealp{2004ApJ...600L..93G}, HST CANDLES; \citealp{2011ApJS..197...35G}), \textsl{Spitzer} (\citealp{2007ApJS..172...86S}, \citealp{2009AJ....138.1261F}), \textsl{IRIS} (\citealp{2014SoPh..289.2733D}), \textsl{Gaia} (\citealp{2016A&A...595A...2G}), \textsl{WISE} (\citealp{Wright2010}), and \textsl{Herschel} (\citealp{2010A&A...518L...1P}). Along with the unprecedented number of new data from space-based facilities, there has been a major improvement in deep and wide surveys with ground-based telescopes. In particular: \textsl{ESO-VLT} (\citealp{2009Msngr.138....4V}), \textsl{Pan STARRS} (\citealp{2002SPIE.4836..154K}), \textsl{Subaru} (\citealp{2007ApJS..172....9T}), and \textsl{CFHT} (\citealp{2007ApJS..172...99C}). Nevertheless, the discrepancy between direct and indirect methods is far larger than expected, and the subject of much debate (\citealp{2019ConPh..60...23M}). At present the discrepancy in the CIB estimates are $\sim75\%$, while in the COB the estimates vary by a factor of 3 to 5. \\

Recently, with the advancement of very high energy facilities, another method has become successful at constraining the EBL by probing the $\gamma$-rays of distant blazars (e.g. \citealp{2008Sci...320.1752M}; \citealp{2011Sci...334...69V}; \citealp{2011MNRAS.410.2556D}; \citealp{2013sf2a.conf..303B}; \citealp{2018MNRAS.476.4187H}; \citealp{2018Sci...362.1031F}). This technique relies on the interaction of high-energy TeV photons from blazars with the COB through $\gamma$-$\gamma$ absorption (\citealp{2017ApJ...850...73D}). Accordingly, this TeV to micron photon coupling allows the use of $\gamma$-rays to probe the optical portion of the EBL over the redshift range covered by the blazar distribution. The advantage of this approach, in comparison with direct estimation, is that it is unaffected by the foregrounds (ZL and Galactic emission), although arguably susceptible to intergalactic magnetic fields. At the present time, the integrated galaxy light (IGL) measurements (EBL estimates in the absence of any diffuse light) agree with the very high-energy (VHE), Dark Cloud, and are mostly consistent with the recent direct estimates from deep space probes (\textsl{New Horizon} and \textsl{Pioneer}, which observe from beyond the ZL and airglow). This consensus motivates us to improve our IGL analysis further by looking to reduce the inherent uncertainty from our previous IGL measurements of $\sim$20\% as reported by \cite{2016ApJ...827..108D}. \\

In \cite{2016ApJ...827..108D}, published data from a variety of sources was assembled spanning from far-UV (FUV) to FIR wavelengths, i.e., covering both the COB and CIB. These data were used to provide compilations of galaxy number-counts from bright to faint magnitudes and integrated to obtain FUV to FIR IGL estimates. The error analysis of that work indicated an uncertainty in the measurements of around 20$\%$ stemming from several sources. In this paper, we look to make two improvements. The first is to replace the bright and intermediate data with a more homogeneous analysis where particular care has been taken to overcome fragmentation and false source detection. The second is an improved error-analysis, in particular an improvement in the sample variance handling and alignment of photometry onto a common filter set. 

Through these improvements we look to improve the optical/NIR IGL uncertainty from 20$\%$ to below 10$\%$, as well as identify the key areas for future improvement to eventually bring the uncertainties down to a few per cent. This paper is structured as follows: In Section \ref{section2}, we summarise the newly processed data from various wide and deep surveys followed by the colour transformations to a common filter system. Section \ref{section3} shows the revised galaxy number-counts and the fitting process. In Section \ref{section4}, we present our analysis for estimating the integrated galactic light of the optical EBL with an improved focus on error analysis. Comparison to VHE analysis and implications for missing diffuse light are discussed in Section \ref{section5}. Finally, we constrain the CSFH using the EBL estimates described in Section \ref{section6}. 
\\

Throughout this paper, all magnitudes are in the AB system and we assume a cosmology with $H_{0}$=70 \rm{km s}$^{-1}$\rm{Mpc}$^{-1}$, $\Omega_{\Lambda}=0.7$, and $\Omega_{M}=0.3$.

%%%%%%%%%%%%%%%%%%%%%
%%% Existing Data %%%
%%%%%%%%%%%%%%%%%%%%%
\section{Number Count Data}
\label{section2}
In \cite{2016ApJ...827..108D} data for the optical/NIR IGL estimates were taken primarily from the GAMA, COSMOS, and HST surveys (HST/ACS/WFC3 + LBT/LBC + ESO VLT Hawk-I). Recently improvements have been made in the analysis of both the GAMA and COSMOS photometry, using the new \textsc{ProFound} source finding code (\citealp{Robotham2018}; \citealp{bellstedt2020galaxy}; Davies et al., in prep). Here we exclusively focus on three contributing datasets: GAMA, DEVILS-COSMOS, and HST, as briefly described below.
% In a future paper, the HST data will also be reprocessed as part of the HST Archival Legacy \textsc{SkySURF} program (PI: R. Windhorst), but are not included here as this work is ongoing and likely to run through 2022.

\subsection{GAMA}
Galaxy And Mass Assembly (GAMA; \citealp{2011MNRAS.413..971D}; \citealp{Liske2015}) is a spectroscopic (98$\%$ completeness) multi-wavelength campaign across $\sim$230 deg$^2$ in the Northern and Southern Galactic caps, centered on R.A.=2h, 9h, 12h, 14.5h, and 23h, with the AAOmega fiber spectrograph on the 3.9 meter Anglo Australian Telescope (AAT) \citep{2013MNRAS.430.2047H}. GAMA includes a panchromatic survey in 20 filters extending from the UV to the FIR \citep{Driver2016}, composed of observations from various wide and deep ground-based (VST KiDS; VISTA VIKING) and space-based (\emph{GALEX}; \emph{WISE}; \emph{Herschel}) facilities as summarised in \cite{Driver2016}. GAMA's main science goal is to quantify the evolution of mass, energy, and structure over the history of the cosmos (\citealp{2018MNRAS.475.2891D}; \citealp{2018yCat..74624336M}; \citealp{2016ASPC..507..269D}). GAMA targets more than 200,000 galaxies to a depth of r $<$ 19.8 magnitude, and presents a panchromatic dataset of galaxies from the low redshift Universe (z $<$ 0.3).\\
Here, we use our reanalysis (\citealp{bellstedt2020galaxy}) of KiDS-DR4 optical-IR (\citealp{2019A&A...625A...2K}) and VISTA VIKING (\citealp{2013Msngr.154...32E}) imaging of the GAMA regions with an effective area of $\sim$217.5 deg$^2$. Previously, GAMA optical photometry as used in \cite{2016ApJ...827..108D}, was based on \textsc{SExtractor} analysis of the SDSS data (see \citealp{2016ApJ...827..108D}), whereas the reanalysis is based on \textsc{ProFound} (\citealp{Robotham2018}). Improvements include the use of dilated segments, watershed deblending, higher-resolution data (KiDS -- SDSS), improved star-galaxy separation, extensive visual inspection, and manual fixing of fragmented galaxies (see \citealp{bellstedt2020galaxy} for full details). These enhancements improve the integrity of the counts at both very bright and intermediate magnitudes, as well as extending the data to fainter limits than was possible with the SDSS data, and in particular to the depths that dominate the IGL measurement.

\subsection{DEVILS-COSMOS}
The Deep Extragalactic VIsible Legacy Survey (DEVILS; \citealp{2018arXiv180605808D}), is a spectroscopic survey on the Anglo-Australian Telescope (AAT), sampling more than $\sim$60,000 galaxies to a depth of $Y < 21.2$ mag. DEVILS covers 6 deg$^2$ of the sky aiming for more than 95$\%$ completeness in three deep fields (COSMOS; \citealp{2007ApJS..172....1S}, ECDFS; \citealp{2006AJ....131.2373V}, XMM-LSS; \citealp{2004JCAP...09..011P}). Here, we adopted the up-to-date photometric catalogue for the D10 (COSMOS) and D02 (CMM-LSS) regions which cover $\sim$4.6 deg$^2$ ($YJHK_s$) of the UltraVISTA and VIDEO datasets (\citealp{2012A&A...544A.156M}; \citealp{2013MNRAS.428.1281J}) centered at RA 10h and DEC 2.2$^{\circ}$ for D10 and RA 2.2h and DEC -4.5$^{\circ}$ for D02  (Davies et al. in prep.). The imaging datasets in \textit{ugrizYJHKs} filters come from the following sources: the \textit{u} band originates from the CFHT, the \textit{griz} bands from the Hyper Suprime Camera Strategic Program (HSC; \citealp{2018PASJ...70S...4A}), and the \textit{YJHKs} bands from either UltraVISTA-DR3 or VIDEO XMM-LSS. For more detail on the photometric catalogue, see the DEVILS photometry description paper (Davies et al., in prep). In the same manner as the updated GAMA photometry, the photometric catalogue has also recently been reassembled using the \textsc{ProFound} source finding analysis package from the FUV to FIR (Davies et al, in prep).

Improvements over previous photometry include those highlighted for GAMA through the use of \textsc{ProFound}, bringing  consistency of methodology, with comparable levels of visual checks and manual fixes.

\subsection{Hubble Space Telescope and other deep data}
For the deepest data, we adopt the deep number counts from \cite{2016ApJ...827..108D}, which were mostly obtained from the HST, Wide Field Camera 3 (WFC3) Early Release Science (ERS) archive (WFC3-ERS; \citealp{2011ApJS..193...27W}) in combination with the HST Ultra-Violet Ultra-Deep fields (UVUDF; \citealp{2013AJ....146..159T}). The data reduction of the panchromatic WFC3 ERS filters are discussed in \cite{2011ApJS..193...27W}. They derived deep field galaxy number counts across the WFC3 ERS filters (F336W, F435W, F606W, F775W, F850LP, F098M, F125W, and F160W) from u band to Y band. The panchromatic photometric counts for the same filters from NUV to NIR in the Hubble Ultra Deep Field (HUDF)  are described in \cite{2015AJ....150...31R}.\\

We note that other HST fields exist, however, we have chosen to exclusively use the ERS and UV-UDF data as they have independent detections in each band, thereby circumventing the colour bias that is introduced through forced aperture photometry. We note that a full reanalysis of the entire ACS WFC3 archive is currently underway, as part of the SkySURF HST Legacy Program (PI: Windhorst), and the inclusion of additional HST fields in IGL EBL analysis will be the subject of a future paper.\\

In the $u$ band we retain the LBT data from \cite{2009A&A...505.1041G} and in the $K_s$ band we also retain the deep data from \cite{2014A&A...570A..11F} using ESO VLT Hawk-I with an exposure time of 13 hours, and reaching a depth of 27.3 AB magnitude (1$\sigma$ limit).\\

Table \ref{datasets_summary} summarises the wavelength coverage and depth of data used in this work.

%%%%%%%%%%%%%%%%%%%%%%%%%%%%%%%%%%%%%%%%%%%%%%%%%%%%%%%%%%%%%%%%%%%%%%%%%%%%%
\begin{table*}
\caption {\label{tab:datasets_summary} A summary of multi-wavelength data used in this analysis. The reference surveys for the analysis are as follows: GAMA (\citealp{bellstedt2020galaxy}); DEVILS (Davies et al, 2020 - in prep); HST ERS Data 3 and UVUDF (\citealp{2011ApJS..193...27W}, \citealp{2015AJ....150...31R}, \citealp{2009A&A...505.1041G}, and \citealp{2014A&A...570A..11F}). The pivot wavelength refers to the GAMA data.} 
\label{datasets_summary}
\begin{tabular}{clllccc} \hline

 EBL & \multicolumn{3}{c}{Survey} & EBL Pivot & \multicolumn{2}{c}{Depth (5$\sigma$ AB)}    \\ 
Filter &  & & & Wavelength  & \\
\cline{2-4} \cline{6-7}
	&  GAMA   & DEVILS   &  HST+   & ($\mu$m) & GAMA   & DEVILS                                             \\    
	\hline \hline
 $u$   &  VLT KiDS   & CFHT(u$^*$)   &   F336W, LBT$^a$ U360$^b$  &   0.3577 &  24.3 & 26.4  \\
 $g$   &  VLT KiDS   & HSC$^c$ $g$   &   F435W  &   0.4744  &  25.1 & 27.3  \\ 
 $r$   &  VLT KiDS   & HSC $r$   &   F606W  &   0.6312  & 24.9 & 26.9  \\
 $i$   &  VLT KiDS   & HSC $i$   &   F775W  &   0.7584  & 23.8 & 26.7 \\
 $Z$   &  VISTA VIKING   & HSC $z$   &   F850LP  &   0.8833  & 23.1 & 26.3 \\
 $Y$   &  VISTA VIKING   & UltraVISTA   &   F098W, F105W  &   1.0224  & 22.3 &  24.7 \\
 $J$   &  VISTA VIKING   & UltraVISTA   &   F125W  &   1.2546 & 22.1 & 24.5 \\
 $H$   &  VISTA VIKING   & UltraVISTA   &   F160W  &   1.6477  & 21.5 & 24.1 \\
 $K_s$   &  VISTA VIKING   & UltraVISTA   &   ESO K$^d$  &   2.1549  & 21.2 & 24.5 \\   \hline
\end{tabular}

 \begin{tablenotes}
    \item[1] a. Large Binocular Telescope (LBT)
    \item[2] b. Large Binocular Camera (LBC)
    \item[3] c. Hyper Suprime Camera (HSC)
    \item[4] d. European Southern Observatory (ESO/VLT/Hawk-I)
  \end{tablenotes}
\end{table*}

\subsection{Colour Transformations}
The datasets described above come from a variety of facilities with aligned but slightly differing filters and facility throughput transmissions. To derive galaxy number counts across consistent bands, we need to implement colour transformations to a common filter set. Here we elect to use the VST $ugri$ and VISTA $ZYJHK_s$ filters as our EBL reference filters. The transformations we need include the DEVILS optical data (CFHT $u^*$ and HSC $griz$ filters), all HST data (F336W, F435W, F606W, F775W, F850LP, F098W, F105W, F125W and F160W), the LBT data ($U360$ filter), and the ESO Hawk-I ($K_s$ filter). For the DEVILS data, we have full access to the flux measurements for each individual galaxy through our segment-matched {\sc ProFound} photometry. Hence, we can convert robustly from native to KiDS filters, by including a colour term to fold in the second-order dependence on the spectral shape. For the HST+ datasets we only have access to the number-count data, {\it which have mostly been determined independently in each band}, and hence can only determine a mean first order filter offset given some reasonable assumption as to the likely mean spectral shape and redshift distribution of the underlying galaxy population (see also the discussion in \citealp{2011ApJS..193...27W}).

To determine our filter transformations, we use the {\sc ProSpect} package (\citealp{Robotham2020}) to generate several thousand galaxy spectra with a uniform redshift distribution from z=0 to z=2, and using the \textsc{ProSpect} default metallicity, dust properties, a Chabrier IMF (\citealp{2003PASP..115..763C}), and a randomly selected star formation history (SFH). We then determine the fluxes for these spectra as observed through our reference filters and through the filters for each of the datasets we wish to transform. We fit for the difference between the measured fluxes in the original and target filters, with a linear dependence on the colour term for DEVILS, and without a colour term for the HST+ data. As part of the fitting process, we determine an estimate of the intrinsic uncertainty in the transformation process, which will of course be higher for the HST+ datasets (because we lack the colour terms). We note that in the future we are looking to reprocess the entire HST archive as part of the {\sc SkySurf} HST Archival program (PI: R. Windhorst), which will allow us, in due course, to factor in the proper HST colour terms. Table~\ref{tab:trans} shows the resulting transformations for the filters used in this work which we now apply.

\begin{table} 
\caption{Filter transformations used to place all data onto the \em{VST/VISTA} filter systems (i.e., \em{ugriZYJHKs}) using our online tool at: \url{https://transformcalc.icrar.org}. \label{tab:trans}}
\begin{tabular}{ll}  \hline
Filter transformation & Intrinsic\\ 
& Uncertainty \\ \hline \hline
$u_{\rm VST} = u^*_{\rm CFHT} + 0.2828 (u^*_{\rm CFHT} - g_{\rm HSC}) - 0.0344$ & 0.056 \\ 
$g_{\rm VST} = g_{\rm HSC} + 0.0331 (g_{\rm HSC} - r_{\rm HSC}) + 0.0163$ & 0.017 \\ 
$r_{\rm VST} = r_{\rm HSC} - 0.0433 (g_{\rm HSC} - r_{\rm HSC}) + 0.0012$ & 0.010\\ 
$i_{\rm VST} = i_{\rm HSC} + 0.0549 (r_{\rm HSC} - i_{\rm HSC}) + 0.0073$ & 0.006 \\ 
$Z_{\rm VISTA} = z_{\rm HSC} + 0.0538 (i_{\rm HSC} - z_{\rm HSC}) + 0.0037$ & 0.007 \\ \hline
$u_{\rm VST} = U360_{\rm LBT} + 0.0038$ & 0.006 \\ 
%$u_{\rm VST} = F300W_{\rm HST} - 0.????$ & 0.??? \\ 
$u_{\rm VST} = F336W_{\rm HST} - 0.113$ & 0.089 \\ 
$g_{\rm VST} = F435W_{\rm HST} - 0.159$ & 0.112 \\ 
$r_{\rm VST} = F606W_{\rm HST} - 0.114$ & 0.070 \\ 
$i_{\rm VST} = F775W_{\rm HST} + 0.031$ & 0.019 \\ 
$Z_{\rm VISTA} = F850LP_{\rm HST} + 0.057$ & 0.049 \\ 
$Y_{\rm VISTA} = F098W_{\rm HST} - 0.076$ & 0.053 \\ 
$Y_{\rm VISTA} = F105W_{\rm HST} + 0.066$ & 0.078 \\ 
$J_{\rm VISTA} = F125W_{\rm HST} - 0.015$ & 0.022 \\ 
$H_{\rm VISTA} = F160W_{\rm HST} - 0.066$ & 0.024 \\ 
$Ks_{\rm VISTA} = Ks_{\rm Hawk-I} + 0.000$ & 0.002 \\ \hline
\end{tabular}
\end{table}

~

~
%These colour transformations are now applied to the DEVILS and HST+ datasets finalising our data selection and corrections.
~
%%%%%%%%%%%%%%%%%%%%%
%%% NUMBER COUNTS %%%
%%%%%%%%%%%%%%%%%%%%%
\section{Results}
\label{section3}
\subsection{Number Counts}
\begin{figure*}
    \centering
    \includegraphics[width=18cm, height=18cm]{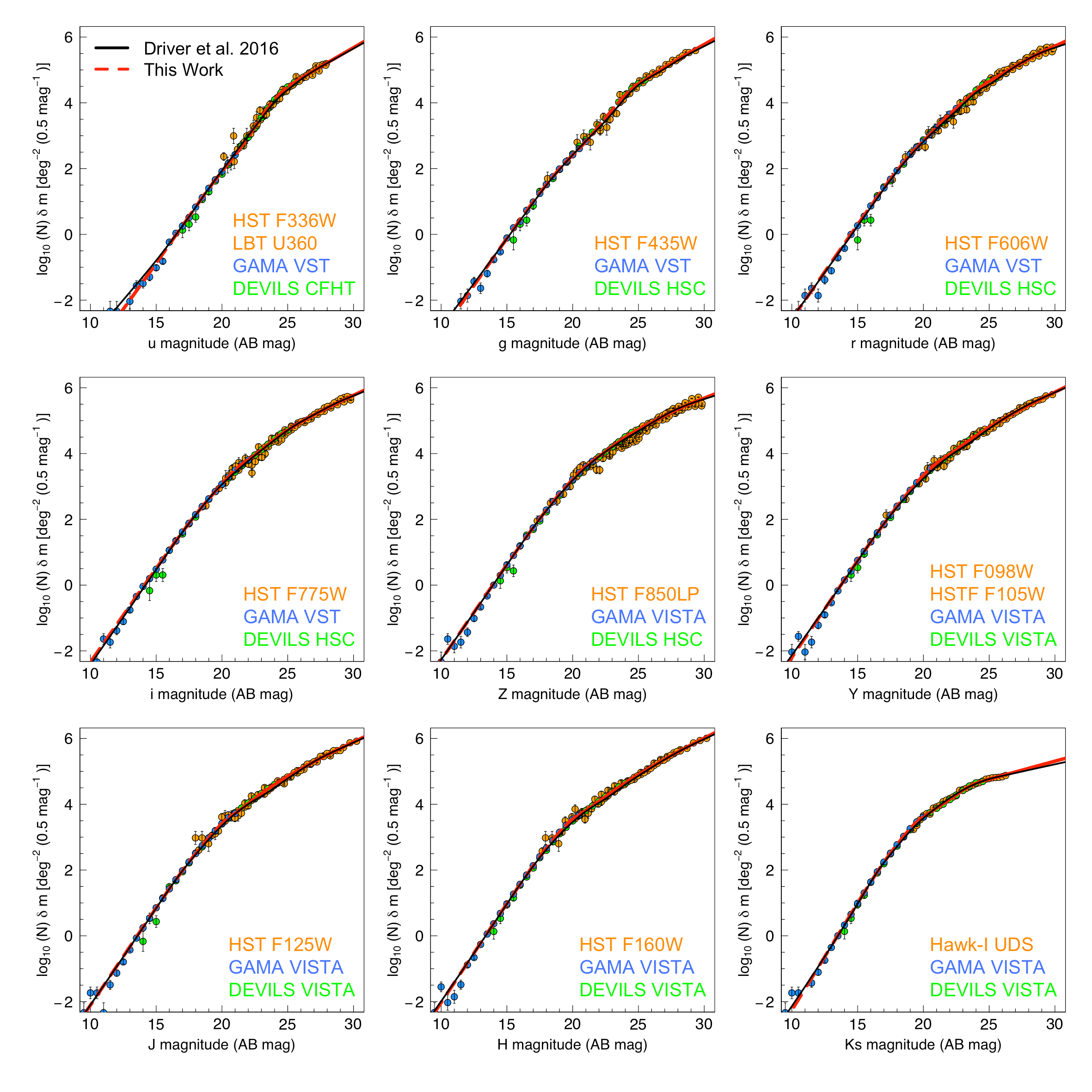}
    \caption{The galaxy number count in each filter for a combination of datasets derived from the GAMA, DEVILS, and the HST surveys in blue, green and orange respectively. We fitted a smooth spline in a red dashed line through all datasets. For a sanity check, we compare to published data (\citealp{2016ApJ...827..108D}). The black solid line shows a fit to their data. The error bars indicate the Poisson error.}
    \label{all_count}
\end{figure*}

\begin{figure*}
    \centering
    \includegraphics[width=18cm, height=18cm]{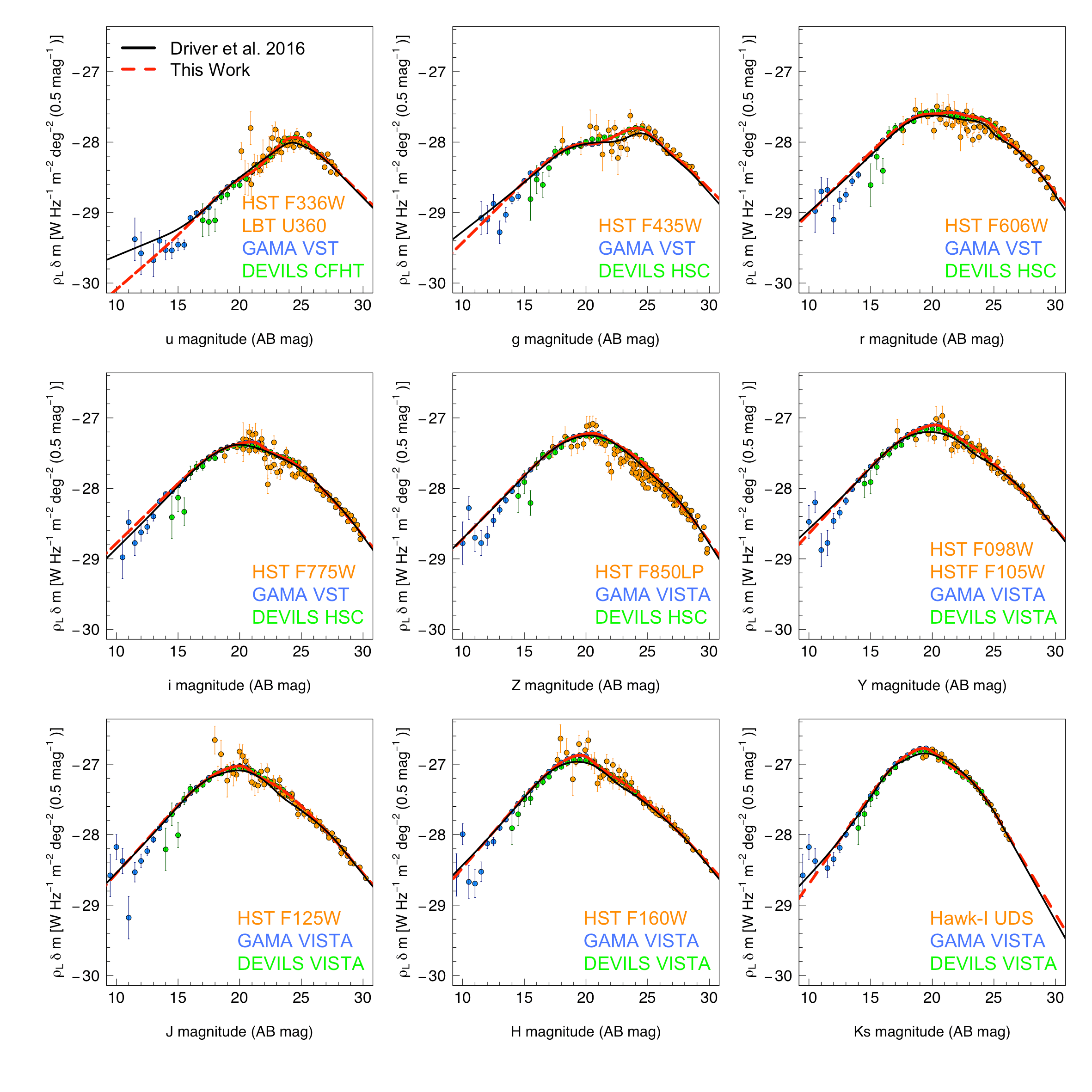}
    \caption{The contribution of each magnitude bin to the energy density. The count has been normalised and scaled for each survey in depth and area.}
    \label{all_density}
\end{figure*}

Figure \ref{all_count} shows our revised galaxy number count compilations in $ugriZYJHK_s$ bands as derived from the deep and wide data (GAMA, DEVILS, and HST+). Figure \ref{all_density} shows the same data, but with a linear term of $d(log_{10}(N))/dm=0.4$ removed to: (a) highlight discrepancies; and (b) highlight the contribution of each magnitude interval to the final EBL. The IGL is simply the integral under these data. In both figures the number counts are in a units of magnitudes per 0.5 mag interval per square degree, and the uncertainties shown are based purely on Poisson statistics. From the observational point of view, the integrated galaxy light derives as below:
    \begin{equation}
        \begin{aligned}
\textrm{IGL} \propto \int N_{m} \times 10^{-0.4 \cdot m} dm
        \end{aligned}
        \label{count_eq}
    \end{equation}
Where \textit{N} is the number of galaxies per magnitude bin between \textit{m} and \textit{m}+$\delta$\textit{m}.

Each dataset is truncated at the faint end at the point at which it becomes inconsistent with deeper data (see Fig. \ref{all_count}). As can be seen, the overlap between the datasets is smooth, with no obvious discontinuities. As a result, the galaxy number counts indicate good agreement in all bands with the largest discrepancies seen in the $u$ and $g$ bands where more robust Galactic extinction corrections may become important (in terms of both spatial distribution and reddening). \cite{2016ApJ...827..108D} discussed possible sources of both systematic and statistical error. These will be discussed in more detail in Section \ref{section4}.\\ 

Figure 2 shows the same plot but now in terms of the differential luminosity density contribution from each magnitude interval. Hence it is the integral under this curve that provides the final EBL measurement in that band. Figure 3 shows the cumulative version of the same plot.

\begin{figure*}
    \centering
    \includegraphics[width=18cm, height=18cm]{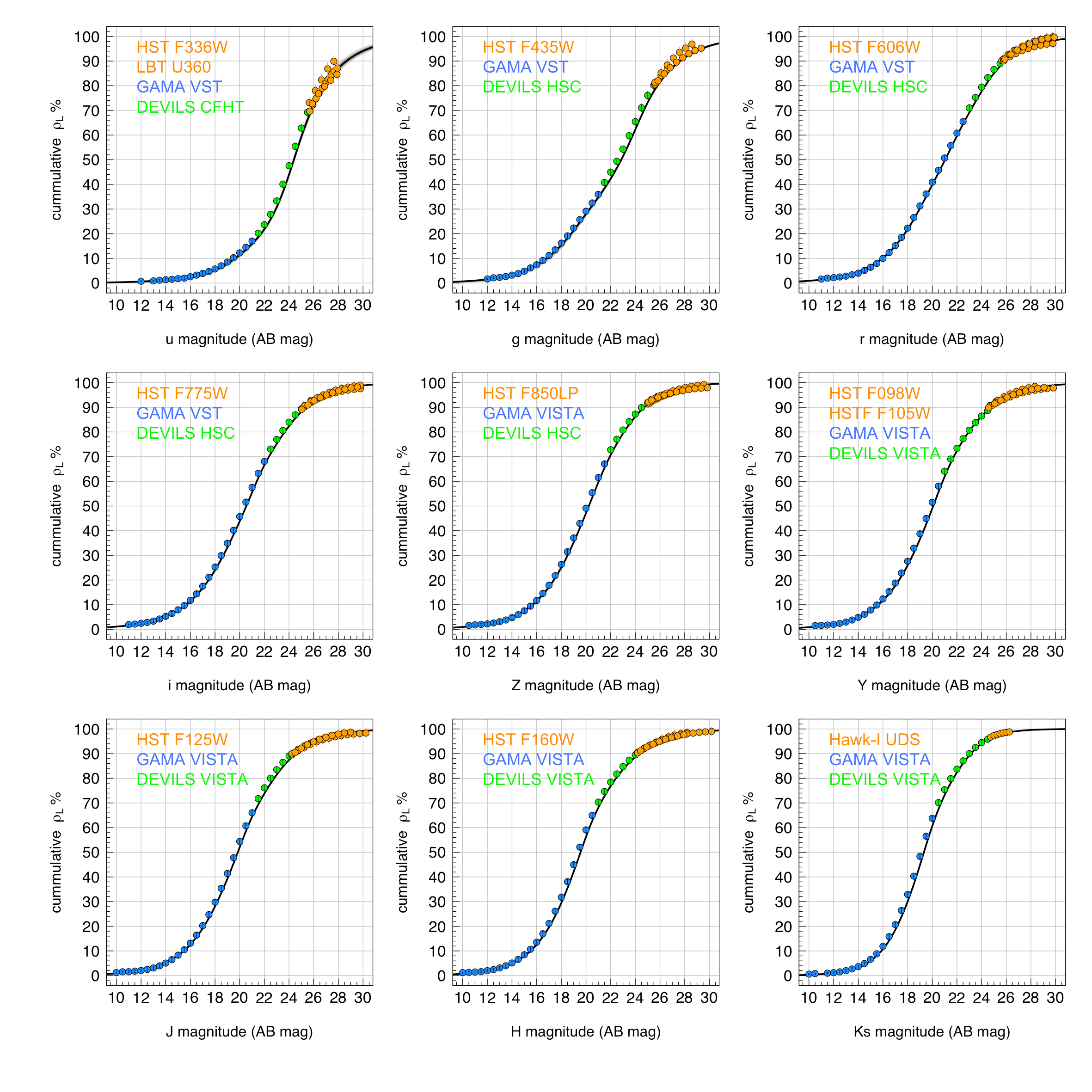}
    \caption{The cumulative contribution function ($\rho_{\rm L}$) to each band.} 
    \label{cumsum}
\end{figure*}
%%%%%%%%%%%%%%%%%%%%%%%%%%%%%%%%%%%%%%%%%%%%%
%%% Extra-galactic Background Light (EBL) %%%
%%%%%%%%%%%%%%%%%%%%%%%%%%%%%%%%%%%%%%%%%%%%%
\subsection{Extragalactic Background Light (EBL)}
In order to estimate the total EBL from discrete sources in each band, we fit a smooth spline with 10 degrees of freedom (df=10) to the combined luminosity density data, inverse weighted by the variance. We then integrate the spline fit from -100 to +100 AB magnitude. Assuming no unexpected behaviour of the spline outside the data range, Fig. \ref{all_density}, shows the contribution of each magnitude bin to the EBL in all filters (as indicated) in the unit of WHz$^{-1}$m$^{-2}$ per 0.5 magnitude bin, and normalised by effective area. As ought to be expected, the contribution to the IGL is limited at both very faint and very bright magnitudes with most of the EBL contribution coming from intermediate magnitudes (and redshift). Following integration we convert to the standard EBL units using Eq. \ref{density_eq}. 

\begin{equation}
    \begin{aligned}
\textsc{EBL} = u \times3282.1\times10^9\times(c/\lambda)\\
    \end{aligned}
    \label{density_eq}
\end{equation}
where $u$ is the energy density, $c$ is the speed of light in $m/s$, and $\lambda$ is the wavelength in meters, and the constant converts the measurements into units of nWm$^{-2}$sr$^{-1}$. Note that the fraction of the IGL in the extrapolated of the spline fit is typically less than a few per cent, see Fig. \ref{cumsum}.
\\

Figure \ref{all_density} shows the luminosity density with the spline fits overlaid (red dashed line) for each band. We compare our best-fitted splines in red dashed lines to the compendium of \cite{2016ApJ...827..108D} shown as solid black lines. Our data are clearly bounded at both ends in all filters and show a well defined peak at intermediate magnitude levels in all bands. 

\section{Error Estimates}
\label{section4}
In this section, we explore error estimation taking into consideration the systematic and random errors for each band. Our systematic errors break down into cosmic variance (CV) and zero-point (ZP) uncertainty, while random errors are due to Poisson error, as well as the uncertainty inherent in the fitting process. The final uncertainty is then the combination of these four separate errors, which we explore in a full Monte Carlo (MC) analysis. Below we discuss each of these individually to address their impact on our final EBL COB measurements.

\subsection{Poisson Error}
\begin{figure}
        \centering
        \includegraphics[scale = 0.4, angle =90]{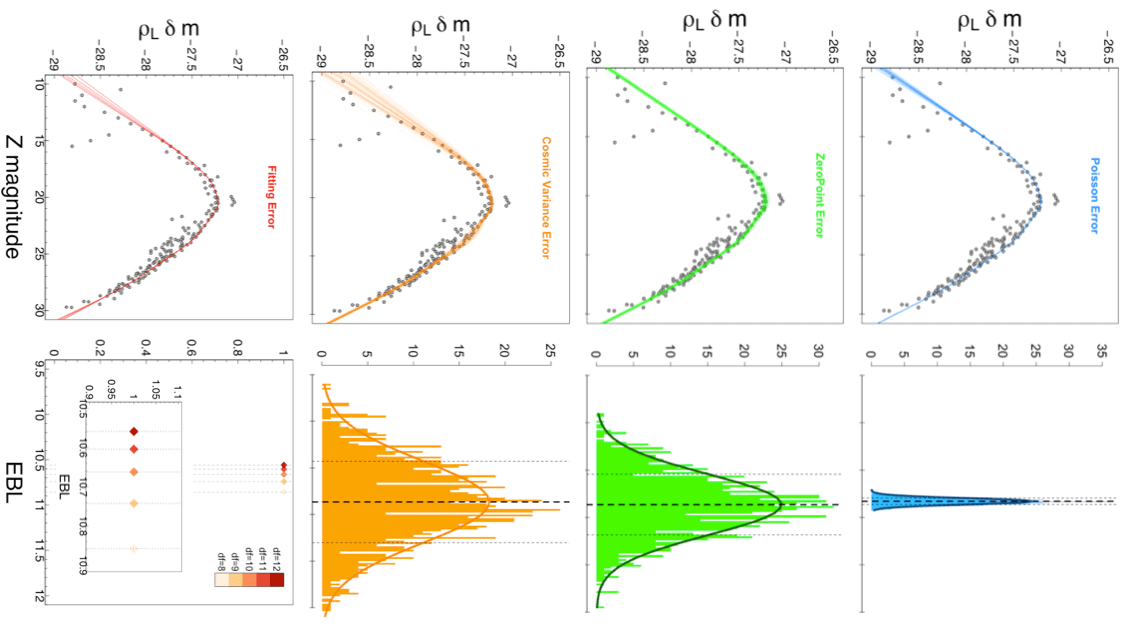}
        \caption{An example of a luminosity density plot including all possible sources of error including the Poisson, CV, ZP, and fitting in blue, orange, green, and red for the Z band, respectively. Left column shows the contribution of each uncertainty to the EBL. Right column shows the distribution of our IGL for each source of error. The mean and standard deviation (1$\sigma$) are shown in black and gray solid lines respectively.}
        \label{error_all}
    \end{figure}

    To estimate the impact of Poisson uncertainty alone, we performed a Monte Carlo analysis with 1001 iterations. Within each iteration, we perturb the galaxy number-counts by their associated density uncertainty error, and repeat our full analysis. For the perturbation of each data point we sample a Gaussian distribution with mean zero and a standard deviation equivalent to the quoted number-count error for that magnitude bin. The top row of Fig. \ref{error_all} shows a range of the MC fits (left panel), and the recovered IGL measurement (right) for 1001 realisations in the Z-band. From this we determine a mean value of 10.658 and a standard deviation of 0.035 i.e., $<$ 1\% error. Table \ref{tab:EBL_tabel} shows this estimate due to Poisson error from $u$ to $K_s$ and in general this error is fairly negligible throughout, as expected given the large sample size.\\
    
    \begin{table*}
        \begin{center}
        \caption{Integrated EBL results considering individual errors. Column 2 shows our IGL measurement in each band. Columns 3 to 6 show the mean offset in $\%$ for the contribution of each uncertainty. Finally, column 7 presents the total error (added in quadrature) as a percentage of the IGL.}
        \label{tab:EBL_tabel}
        \begin{tabular}{l l l l l l l}
        \hline
        \hline
        Filter &  EBL & Poisson Error & Cosmic Variance & Zero-Point & Fitting Error & Total \\
         &$(nW/m^{2}sr^{-1})$&($\%$)& Error ($\%$)& Error ($\%$)&($\%$)&($\%$ of IGL)\\
        \hline
        \hline
        u & 4.13 & 0.38 & 2.54 & 6.36 & 0.35 & 6.87 \\
        g & 5.76 & 0.49 & 3.29 & 2.07 & 0.92 & 4.02 \\
        r & 8.11 & 0.37 & 3.55 & 1.44 & 1.34 & 4.08 \\
        i & 9.94 & 0.35 & 3.98 & 1.67 & 0.99 & 4.44 \\
        Z & 10.71 & 0.30 & 3.98 & 3.16 & 1.06 & 5.20 \\
        Y & 11.58 & 0.32 & 4.11 & 1.62 & 0.88 & 4.52 \\
        J & 11.22 & 0.29 & 4.25 & 2.35 & 0.75 & 4.92 \\
        H & 11.17 & 0.26 & 4.38 & 1.63 & 0.65 & 4.73 \\
        Ks & 9.42 & 0.19 & 4.12 & 3.09 & 0.74 & 5.21 \\
        \hline
        \end{tabular}
        \end{center}
    \end{table*}

\subsection{Zero-Point Error}
\label{zp_error}
\begin{figure*}

    \includegraphics[width=8.5cm]{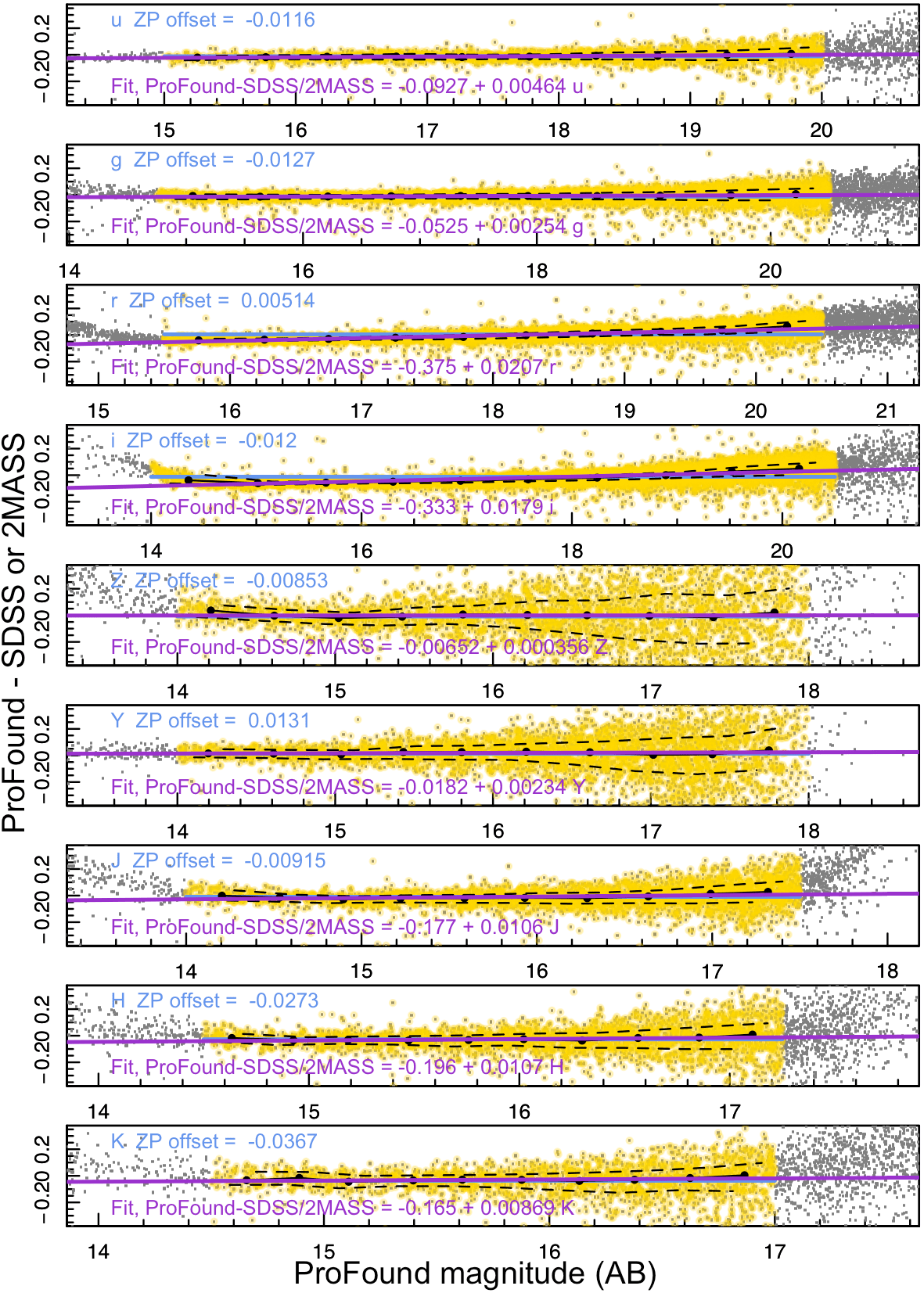}
    \includegraphics[width=8.5cm]{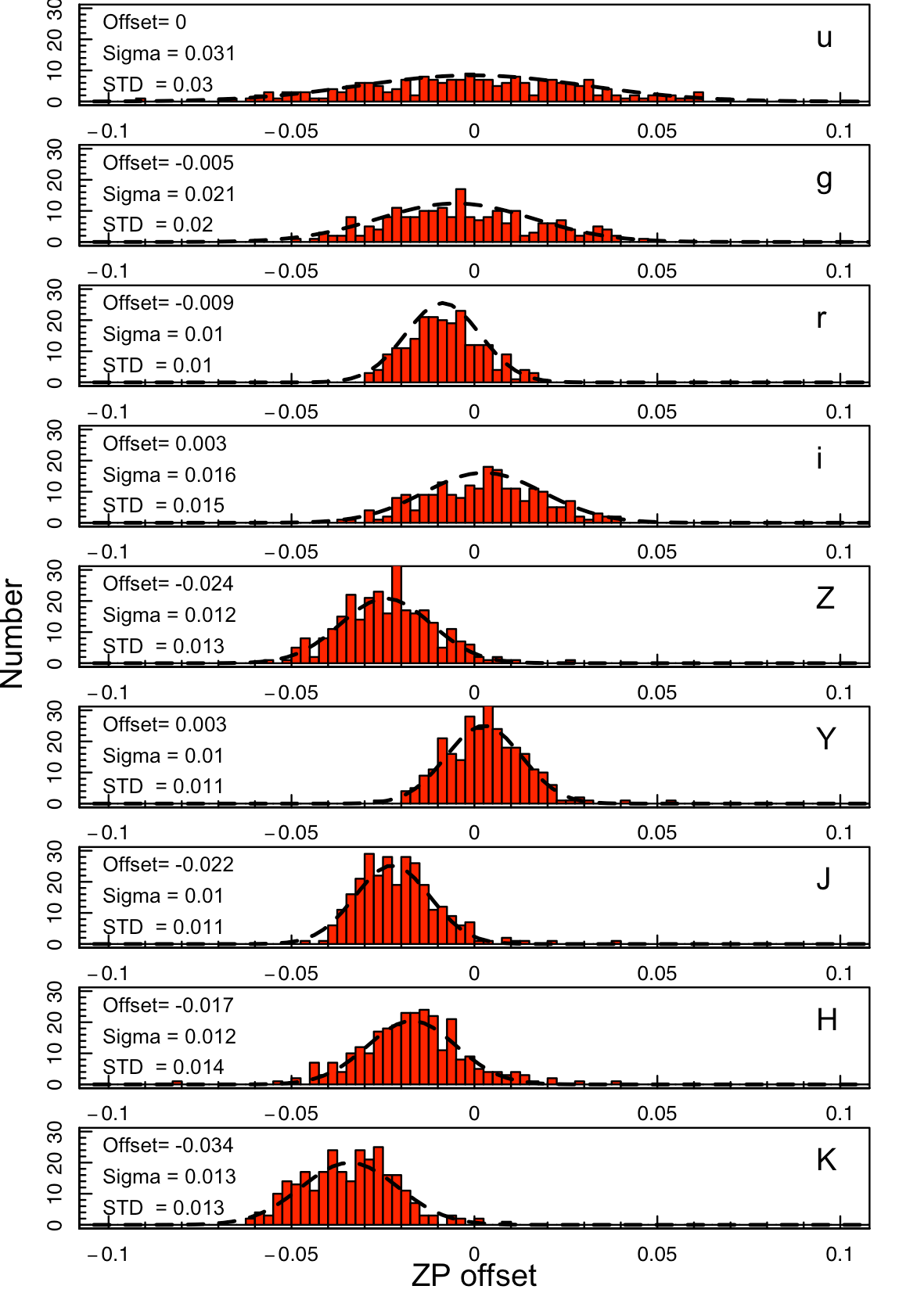}
    
    \caption{Left panel: A direct comparison of the GAMA \textsc{ProFound} photometry to either SDSS ($ugriZ$) or 2MASS ($YJHK_s$) data, using objects classified as stars. The region we use for determining offsets for each {\sc TILE} and for each filter are shown in yellow, with the offsets and line fits for each sample shown in blue or mauve respectively. Right panel: the compendium of zero-point offsets for all 220 {\sc TILES} with the $\sigma$ of the Gaussian fit and the standard deviation shown on each panel.}
    \label{fig:zp}
\end{figure*}

Errors in the absolute zero-point calibration will also affect the number-counts in a systematic manner for a given dataset, but randomly between the datasets, and randomly between filters. Here we estimate the zero-point errors in the GAMA/KiDS data by comparing
the KiDS/VIKING photometry to the SDSS/2MASS photometry, incorporating appropriate filter conversions as provided by \cite{2019A&A...625A...2K} for KiDS (only relevant for $u$ and $r$ band, see their figure 5 and eq~5), and \cite{2018MNRAS.474.5459G} for VISTA (in $ZYJHK_s$ bands, see their eqs 5---9 and also Appendices A and C). In effect, we are attempting to provide an independent albeit basic calibration check of the KiDS and VIKING data, compared to the more sophisticated calibration conducted by the KiDS and VISTA CASU (Cambridge Astronomy Survey Unit) teams. The zero-point verification analysis is done on a square-degree by square degree (i.e., {\sc TILE}) basis, and we exclusively use stars in a restricted magnitude range, where bright stars are not saturated in KiDS/VIKING, and faint stars are not dominated by sky noise for SDSS/2MASS. Left panel in Fig.~\ref{fig:zp} shows an example comparison for the $ugriZYJHK_s$ filters for a single {\sc TILE} centred at RA=8.5h and Dec=$-0.5^o $, the yellow data points highlight the magnitude range used, while the grey data points showing the full range of the comparison. From the yellow data points, we determine a simple offset and a linear fit for each filter and for each {\sc TILE}, as indicated by the cyan and mauve lines respectively. In general the fits are good, showing relatively little magnitude dependence, and hence we use the simple offsets as our indicator of the zero-point offset for each {\sc TILE}. Figure ~\ref{fig:zp} (right) shows the histogram of these zero-point offsets for the 220 {\sc TILES} that make up the GAMA KiDS/VIKING dataset reflecting a distribution of zero-point uncertainty. For each filter, we determine the median absolute zero-point offset, and the 1-$\sigma$ zero-point uncertainties from a simple Gaussian fit, and the empirical standard deviations (as indicated in the Figure panels). We also note that the KiDS team identified a Galactic Latitude dependency of --0.02 to 0.1 mag in the $u$ band, potentially explaining the much broader spread that we see here. We also note that the VISTA zero-points evolved from VISTA reduction 1.3 to 1.5 with implied zero-point changes of --0.03, 0.018, --0.0200, 0.0067 and 0.0106, for the five VISTA bandpasses ($ZYJHK_s$) respectively (see \citealp{2018MNRAS.474.5459G}, Appendix D). We therefore elect to adopt worst case scenario for the zero-point errors by combing all of the uncertainties in quadrature (i.e., literature offset + median offset + (1-$\sigma$) spread + intrinsic uncertainty from Table \ref{tab:trans}) to obtain ultra-conservative zero-point uncertainties of: 0.11, 0.022, 0.011, 0.018, 0.041, 0.021, 0.032, 0.022 and 0.037 for the GAMA $ugriZYJHK_s$ respectively.
    
For the DEVILS and HST+ datasets we adopt the zero-point uncertainties as the maximum of either those found for GAMA, or an error floor of 0.03 mag for DEVILS (i.e., that adopted in \citealp{2018MNRAS.475.2891D}) or HST+. In addition we also fold into the DEVILS and HST+ zero-point errors the intrinsic uncertainty from the filter transformations shown in Table~\ref{tab:trans}. Technically, this overestimates the likely colour transformation error (as it will be random not systematic), but this is a convenient place to incorporate it that errs on the side of caution. Table~\ref{table:zp} shows the final adopted zero-point errors for each dataset and each band. 
    
\begin{table*}
    \caption{Adopted zero-point uncertainties for each dataset and each filter. \label{table:zp}}
    \begin{tabular}{c|c|c|c|c|c|c|c|c|c} \hline
   Dataset  & $u$ & $g$ & $r$ & $i$ & $Z$ & $Y$ & $J$ & $H$ & $K_s$ \\ \hline \hline    
   GAMA    & 0.11 & 0.02 & 0.01 & 0.02 & 0.04 & 0.02 & 0.03 & 0.02 & 0.04 \\
   DEVILS & 0.11 & 0.03 & 0.03 & 0.03 & 0.04 & 0.03 & 0.03 & 0.03 & 0.04 \\
   HST+ & 0.11 & 0.08 & 0.13 & 0.06 & 0.04 & 0.04 & 0.07 & 0.03 & 0.04 \\ \hline
    \end{tabular}
\end{table*}
    
With conservative zero-point errors in hand, we can now Monte Carlo simulate the impact of the zero-point error by perturbing the magnitudes for each of our three datasets (GAMA, DEVILS, and HST) in a systematic manner, but independently for each dataset and for each filter. To do this, we draw a random value for each dataset and filter from a Gaussian distribution with a mean of zero and a standard-deviation defined by the relevant zero-point uncertainty (Tab~\ref{table:zp}), and systematically adjust the magnitude bins of each survey by this amount for that filter. Finally, we refit our splines and re-derive our optical/NIR IGL measurements 1001 times to build up a distribution of the impact of the zero-point errors. The results are shown in Table \ref{tab:EBL_tabel} and Fig. \ref{error_all}. 
    
\subsection{Cosmic Variance Error}
    Pragmatically, cosmic variance can be considered as a systematic uncertainty inherent in observational estimates of the volume of the extragalactic sky due to large-scale density fluctuations. This error is quite significant in most galaxy surveys, and is often the dominant error in measuring the IGL (especially in deep galaxy surveys).
    
\begin{figure}
        \centering
        \includegraphics[scale = 0.33]{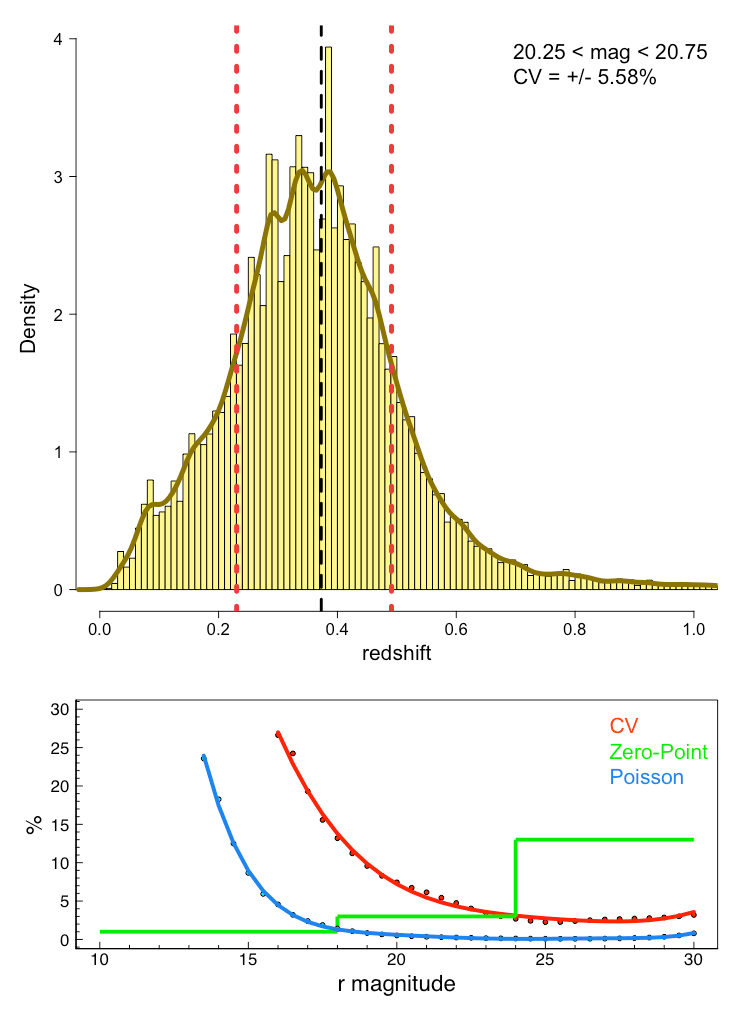}
        \caption{Top panel: Cosmic variance density distribution for a sample of a magnitude bin in the $r$ band. The dashed black line denotes the mean redshift. Upper/lower quantiles are displayed in dashed red lines. Bottom panel: The cosmic variance compared to the zero-point and Poisson uncertainties.}
        \label{fig:CV}
    \end{figure}

    For the GAMA survey, we obtained a mock catalogue using the \textsc{SHARK} semi-analytic model of galaxy formation and evolution (\citealp{2019MNRAS.489.4196L}; \citealp{2020arXiv200311258B}) used to construct light-cones for the GAMA regions. To estimate our CV error, we can determine the effective volume from the 16\% and 84\% quantiles of the predicted redshift distribution, $n(z)$, for the magnitude slice in question, and evaluate equation 3 from \cite{Driver2010}. We use an effective sky area of 60 deg$^{2}$, redshift ranges of the comoving cone, and set the number of separated regions to 4 for each survey to obtain the CV error. Top panel in Fig. \ref{fig:CV} shows an example redshift distribution for the magnitude range of 20.5$\pm$0.25 in the $r$ band. Fig. \ref{fig:CV} (bottom panel) also highlights the variation of the cosmic variance as a percentage (solid red line) compared to the zero-point (solid green line) and Poisson (solid blue line) uncertainties.   
    
    For the DEVILS/D10+D02 and HST surveys, we also adopted the cosmic variance uncertainty following eq 3 of \cite{Driver2010} for the appropriate sky area for the particular survey, and adopting an approximate redshift range. To determine the approximate redshift ranges to use for the DEVILS/D10 + D02 and HST surveys, we sliced a mock galaxy catalogue generated by the SHARK team into the 16\% and 84\% quantile ranges independently for each magnitude bin and each bandpass. For our three datasets, we then calculated CV fluctuations for each magnitude bin, and perturb the galaxy counts in the magnitude direction using this value for each filter and independently for each dataset. We perform a Monte Carlo simulation with 1001 iterations and, consequently, rederived the IGL EBL. Throughout the CV analysis, the procedure was forced to start from the same random seed to ensure that the CV error is correlated across the filters. Table \ref{tab:EBL_tabel} shows the error from the CV.     
    
\subsection{Fitting Error}
    To calculate the IGL, we performed a spline fit using an arbitrarily selected 10 degrees of freedom. Figure \ref{error_all} (lower) shows the shift in the $Z$ band IGL value if we use a spline of order 8, 9, 10, 11 or 12. On the whole, the shift is marginal, as compared to that from the Poisson error (see top panel in Fig. \ref{error_all}), and far less than the effect of either cosmic variance or the zero-point errors.
  
    \begin{figure}
        \centering
        \includegraphics[scale = 0.23]{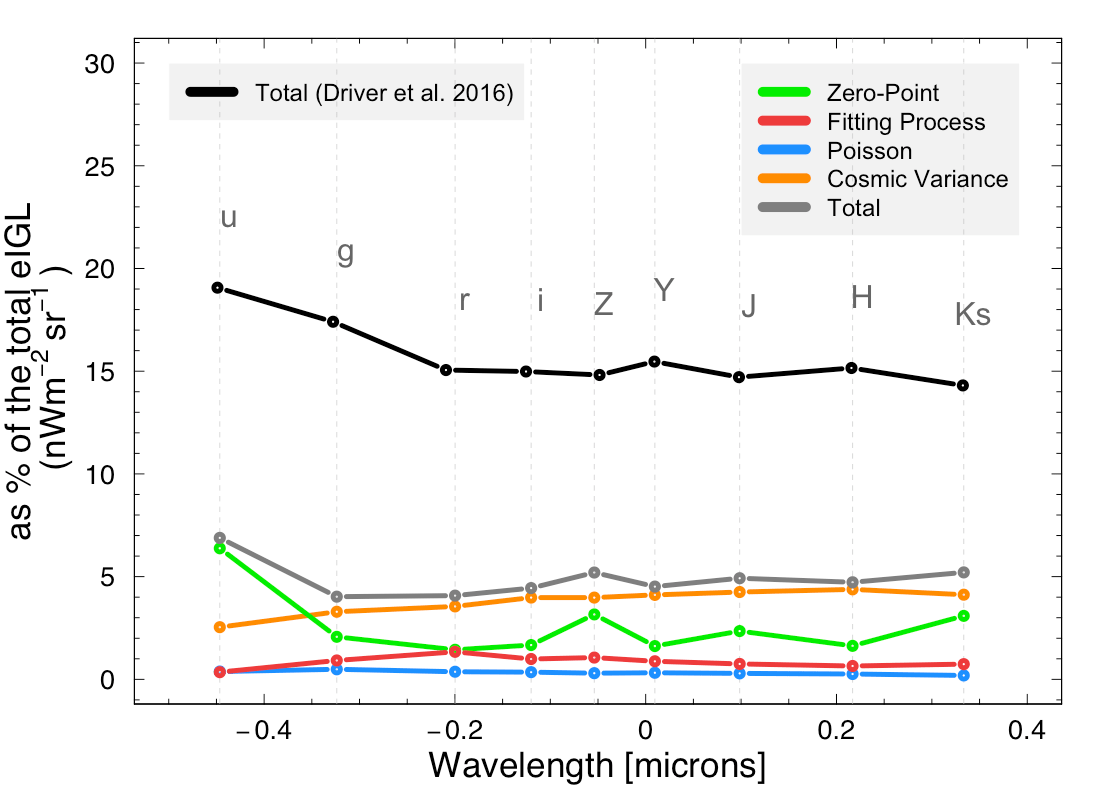}
        \caption{A contribution of each systematic and random uncertainties to the IGL for each filter. The contribution of the cosmic variance to the IGL is dominant across the optical and NIR filters. Random errors have an insignificant contribution to the IGL estimation. Our total error (in gray line) has been compared to ~\protect\cite{2016ApJ...827..108D} marked with black line.}
        \label{error_budget}
    \end{figure}

\subsection{Final error}
Table \ref{tab:EBL_tabel} summarizes the integrated EBL results showing the individual errors. Figure \ref{error_budget} shows the contribution of each error as a percentage of the optical/NIR IGL in a single graph. We show the photometric, spline fitting, Poisson and cosmic variance error in green, red, blue and orange lines, respectively. As can be seen, the systematic cosmic variance and zero-point errors dominates for all filters, while the contribution of random errors (fitting and Poisson) is negligible. The solid line also highlights the total error budget for all optical and near-IR filters ($ugriZYJHK_s$) comparing to the previous result from \cite{2016ApJ...827..108D}; shown as a black line. \\

We finally combine the errors in quadrature, to produce our final IGL error estimates. However we note, that the errors are a mixture of both random and systematic errors. Table \ref{tab:EBL_tabel} (last column) shows our final error estimate as a percentage of the IGL value. As a consequence of the revised data, colour transformations and revised error analysis, the IGL error, compared to the earlier results of \cite{2016ApJ...827..108D}, is reduced from $\sim$20\% to the mean error of $\sim$8\%. \\

%%%%%%%%%%%%%%%%%%
%%% COB RESULT %%%
%%%%%%%%%%%%%%%%%%
\section{Cosmic Optical Background Result}
\label{section5}
\subsection{Comparison to previous IGL Data}
\begin{figure}
    \centering
    \includegraphics[scale = 0.33]{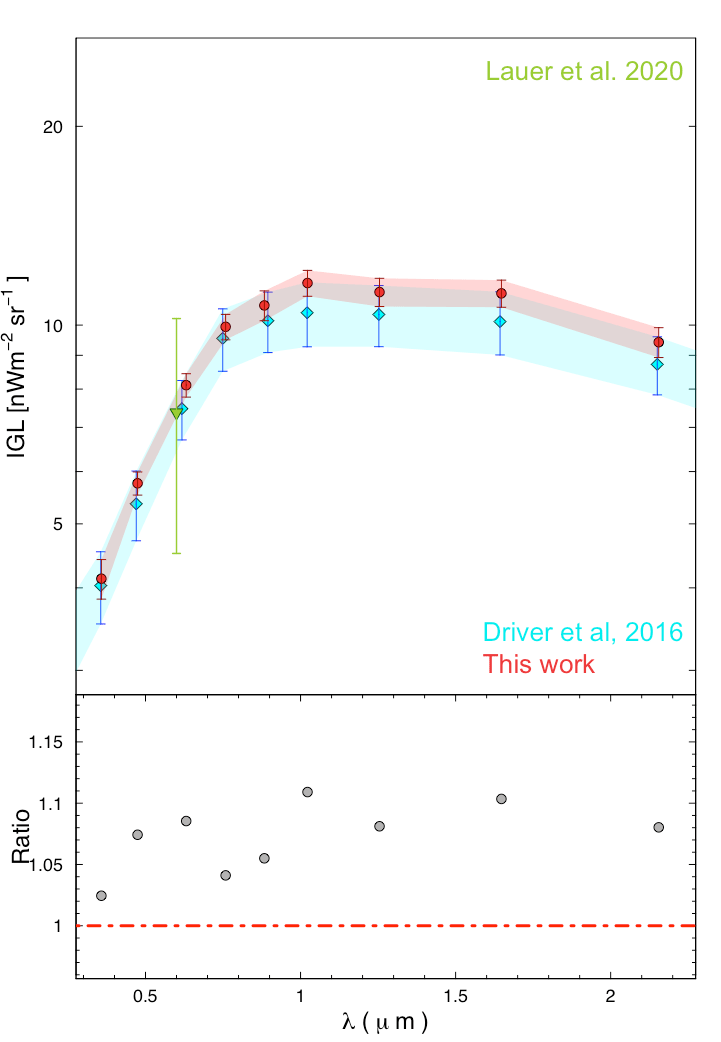}
    \caption{Top panel: The IGL as a function of wavelength estimated in ~\protect\cite{2016ApJ...827..108D} and this work marked with cyan diamonds, and red circles, respectively. Shaded areas correspond to the total error. For comparison, we show the IGL estimate from ~\protect\cite{2020arXiv201103052L} as a lower limit. As can be seen, their estimate is consistent with our measurement. Bottom panel: The ratio of IGL measured in this work to the IGL measured by ~\protect\cite{2016ApJ...827..108D}. The red dot-dashed line corresponds to the equality line between the two measurements.}
    \label{error_ratio}
\end{figure}

The top panel in Fig. \ref{error_ratio} shows the comparison between our new IGL measurements and the results from \cite{2016ApJ...827..108D}, who used essentially the same methodology, but which we have improved by the replacement of SDSS data with VST KiDS data, a full reanalysis of the core GAMA and COSMOS/DEVILS number-counts data, as well as an improved error analysis. We also include the very recent data from the New Horizons LORRI (\citealp{2020arXiv201103052L}) using the \cite{2017NatCo...815003Z} (green triangle) estimates which were obtained at an even greater distance of $>$ 40 a.u. from the Sun. The data is obtained at the LORRI pivot wavelength of 0.608 $\mu$m at high galactic latitude after subtracting estimates of the outer Kuiper belt light and DGL. The \cite{2020arXiv201103052L} IGL estimate is measured as 7.37 $\pm$ 0.81 $\pm$ 2.05 (random and systematic errors respectively) nWm$^{-2}{\rm sr}^{-1}$, which looks consistent with an interpolation of our IGL points at 0.6 microns.\\
The ratio of our revised measurement of the IGL for each filter, to that from \cite{2016ApJ...827..108D} is shown in the lower panel of Fig. \ref{error_ratio}, and highlights the changes resulting from the current analysis. As the areas covered by the contributing surveys are essentially identical the reduction in error comes mainly from the improvement in our modeling of the cosmic variance, which now relies on the SHARK galaxy formation simulations (see \citealp{2019MNRAS.489.4196L}), and improved colour corrections (see \citealp{Robotham2020}). In all wavebands, we see modest increase in the recovered IGL value of approximately 5-15 per cent (see lower panel of Fig.~\ref{error_ratio}). While in all bands the change is within the errors, it is systematic across the filters. We believe this is due to a combination of {\sc ProFound} recovering more accurate total fluxes than the previous source detection algorithms (see Fig. 14 in \citealp{Robotham2018}), the improved depth of KiDS over SDSS, improved star-galaxy separation (see \citealp{bellstedt2020galaxy}), and to a lesser extent more accurate accounting of the masked area, i.e., that lost near bright foreground stars. We note that this increase in the IGL measurements brings into line the IGL and the VHE data, available at that time, in which \cite{2016ApJ...827..108D} noted a 10-20\% discrepancy. At the time this was tentatively attributed to the possibility of stripped mass (e.g., ICL and Intra-Halo Light) and/or missing populations. See also all discussion in \cite{2018PASP..130f4102A}. They found $\leq$ 10-20\% of light to surface brightness $\sim$32 mag.

\subsection{Comparison to Very High Energy (VHE) constraints}
 As indicated in the introduction, VHE studies estimated the EBL intensity through the production of electron-positron pairs as a by-product of the interaction of high energy photons, emanating from the distant blazars, with micron wavelength photons in the EBL. This interaction results in the decrement of the expected power-law distribution of high energy photons with 100GeV blazar photons preferentially interacting with micron wavelength EBL photons. At present, most VHE applications require an input EBL model curve and constrain the normalisation of the curve. Figure \ref{ebl_total_plot} presents the most recent results from \cite{2018Sci...362.1031F}, who studied 759 active galaxies as observed via the Fermi Large Area Telescope (LAT). In Fig. \ref{ebl_total_plot} we compare our new IGL measurements to published VHE COB measurements (\citealp{2015ApJ...812...60B}; \citealp{2016A&A...590A..24A}; \citealp{2018MNRAS.476.4187H}; \citealp{2018Sci...362.1031F}). This latest result shows full consistency with our optical/NIR IGL data, however, we note that an improved comparison could be made by a re-analysis of the VHE data which adopts an EBL model fitted to our IGL data (see Sec. \ref{section6}).
 
 Current attempts, within the VHE community, are now ongoing to not only constrain the EBL normalisation, but also the shape of the EBL COB SED (see for example \citealp{2019ApJ...885..150A}). Recently, the VERITAS team (\citealp{2015ApJ...812...60B}) demonstrated this by detecting and measuring the spectra of 14 blazars at very high energy ($>$ 100 Gev) in order to enable a full reconstruction of the SED of the EBL. However, at this stage the analysis errors are too large to be useful. \\
 
The shaded area obtained by the VHE COB measurement also depicts the range of the allowed EBL intensity in the optical/near-IR filters (from 0.24$\mu$m to 4.25$\mu$m for MAGIC and extending to 10.4$\mu$m for H.E.S.S). At this stage it can only really be stated that the VHE data are fully consistent with our IGL measurements in the $u-K_s$ range without the need to include any significant additional source of diffuse light. 

As the errors in both the optical/NIR IGL and VHE COB data improve to the few percent level, the comparison between these complementary methods should provide interesting constraints on the presence, or lack of non-bound optical/NIR emission.

\subsection{Comparison to direct EBL constraints from space platforms}
\begin{figure*}
    \centering
    \includegraphics[scale = 0.32]{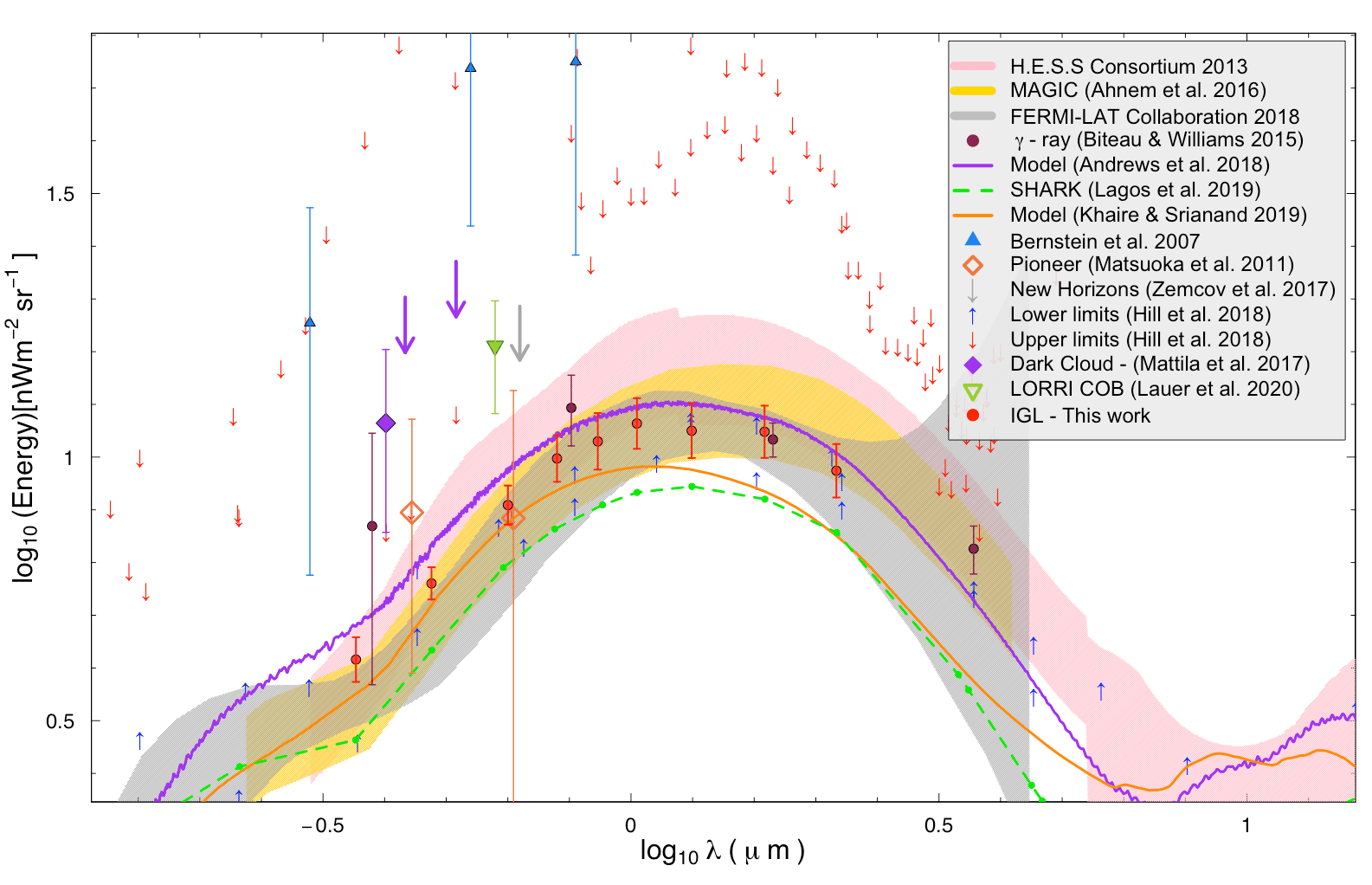}
    
    \caption{Our measurement of the IGL based on galaxy number counts from wide and deep surveys (GAMA+DEVILS+HST) in nine multi-wavelengths from $u$ to $K_s$ band (red circles). We compared our IGL to literature measurements taken from various techniques. Direct estimates:~(\protect\citealp{Bernstein_2007}) in blue triangles; New Horizons~(\protect\citealp{2017NatCo...815003Z}) as an upper limit in gray; Pioneer 10$/$11~(\protect\citealp{2011ApJ...736..119M}) in orange diamonds; recent estimates from the New Horizons LORRI ~(\protect\citealp{2020arXiv201103052L}) in green triangle. VHE estimates:~(\protect\citealp{2008Sci...320.1752M}) in yellow shades;~(\protect\citealp{2018MNRAS.476.4187H}) in pink shades;~(\protect\citealp{2018Sci...362.1031F}) in gray shades; $\gamma$-ray observation from~(\protect\citealp{2015ApJ...812...60B}) in violet circles. Dark Cloud:~(\protect\citealp{2017MNRAS.470.2133M}) and upper limits in purple. We also compared the EBL estimation obtained from the models: SHARK~(\protect\citealp{2019MNRAS.489.4196L}) green dashed line; ~(\protect\citealp{2019MNRAS.484.4174K}) orange line; GAMA~(\protect\citealp{2018MNRAS.474..898A}) purple line. Finally, we added a data assembled by~\protect\cite{2018ApSpe..72..663H} from radio to $\gamma$-ray wavelengths as an upper and lower limit following by the red and blue arrows respectively.}
    
    \label{ebl_total_plot}
\end{figure*}

Figure \ref{ebl_total_plot} shows a comprehensive compilation of nearby Solar system measurements of the COB. These mainly stem from direct background measurements of HST data, and are often shown as either upper limits (\citealp{Bernstein_2007}) or sometimes shown as tentative measurements (\citealp{2018ApSpe..72..663H}). The open orange diamonds and gray square show direct estimates from the Pioneer 10/11 (\citealp{2011ApJ...736..119M}) and New Horizons (\citealp{2017NatCo...815003Z}) spacecraft at distances $>$ 4.5 a.u. respectively. Pioneer 10/11 reported measurement of the reanalysed EBL beyond 4.5 a.u. as $7.9_{+4.0}^{-4.0}$ nWm$^{-2}{\rm sr}^{-1}$ and $7.7_{+5.8}^{-5.8}$ nWm$^{-2}{\rm sr}^{-1}$ at 0.44 and 0.64 $\mu$m respectively. New Horizon also reported the EBL as $4.7_{+7.3}^{-7.3}$ nWm$^{-2}{\rm sr}^{-1}$ at 0.66 $\mu$m. Both of these measurements are sufficiently far from the Sun to be considered near free of Zodiacal light contamination. 

\label{EBL}

As clearly shown in Fig. \ref{ebl_total_plot}, our optical/NIR IGL measurements are inconsistent with the interpretation of the {\sc HST}-based direct estimates as tentative measurements, but consistent as upper limits. As outlined in \cite{Bernstein_2007}, the key issue is robust subtraction of bright foregrounds such as the diffuse Galactic light and the Zodiacal emission. The discrepancy is less prominent between the direct measurements from deep-space probes such as the Pioneer 10/11 or the New Horizon mission but to a lesser extent. It would therefore seem that the likely issue, at least in comparison to the HST data, is with the subtraction of the Zodiacal light, which drops substantially as one moves outward in the Solar system, or potentially due to additional airglow contamination from the upper atmosphere. For completeness, we also include recent result from the dark cloud method (purple diamond; \citealp{2017MNRAS.470.2152M}), which excluded bright galaxies $<$ $\sim$22 mag and resulted in $11.6_{+4.4}^{-4.4}$ nWm$^{-2}{\rm sr}^{-1}$ at 400 nm as an upper limit.\\

Hence some small tension does still remain between the deep space probe data, the dark cloud data, and the IGL data which as argued by \cite{2020arXiv201103052L} may leave some room for a diffuse COB component of unknown origin. One argument against this, is the strong agreement between the IGL data which is sensitive to the discrete COB, and the VHE data (discussed earlier) which is sensitive to both the discrete and diffuse COB. Hence we are in the intriguing situation where the IGL and VHE are fully consistent, but at low-level tension with the deep space platform and dark cloud data. An obvious solution is that the additional radiation being detected by the space-platform data may be associated with either the outer Solar System, additional Diffuse Galactic Light, or the outer Milky-Way dark-matter Halo, i.e., local. Resolving this discrepancy will likely require improvements in all three methods.

Finally, in Fig. \ref{ebl_total_plot} we also show three recent models along with our data points. \textsc{SHARK} (\citealp{2019MNRAS.489.4196L}; dashed green line) is a semi-analytical model of galaxy formation and evolution (SAM) extending from FUV to FIR, the model of \citep[][solid orange line]{2019MNRAS.484.4174K} is a synthesis model of the EBL from FIR to high energy $\gamma$-rays, and \cite{2018MNRAS.474..898A} (solid purple line) presents a phenomenological model of the cosmic spectral energy distribution (CSED) based on simple forward modelling from an input cosmic star formation history (CSFH). While all the models show good consistency they mostly lie outside of the error ranges of our data heralding the prospect of using our data to now test the models more stringently as our errors continue to improve.\\     
\\
In this work, we have shown that the optical/NIR IGL uncertainty has been improved (see Fig.\ref{error_budget}) in the optical regime from 20\% (\citealp{2016ApJ...827..108D}) to below 10\%.
Figure ~\ref{fig:CV} highlights the dominant error across the magnitude range in the $r$ band. This shows that in order to further improve IGL measurements we need wider area data at brighter magnitudes as well as improved calibration, including filter conversions, at fainter magnitudes. Both of these will be satisfied through the WAVES and HST \textsc{SkySurf} projects. The former will reprocess the entire VISTA VIKING/VST KiDS region increasing the wide area coverage from 230 deg$^2$ to 1300 deg$^2$, while the \textsc{SkySurf} project will include a complete reanalysis of HST data using \textsc{ProFound} with the opportunity to perform more robust colour corrections on a galaxy-by-galaxy basis. Both of these improvements have the potential to reduce the IGL uncertainty to $<$3\%. 

In due course the Rubin Observatory, EUCLID, the Roman Space Telescope, and JWST will potentially provide wider and deeper data to reach a potential accuracy of $<1$\% within 10-20 years. In addition, the suitability of recent surveys such as Pan-STARRS, DES, the DESI Legacy Survey, and HyperSuprimeCam at Subaru should provide a promising pathway forward as our measurements improve.
 
%###################################################################################################
%################                   COSMIC STAR FORMATION HISTORY               ####################
%###################################################################################################

\section{Modelling the EBL}
\label{section6}
 As outlined in the introduction, the COB (and CIB) is the radiation by-product of cosmic star-formation over all time, and as such is directly predictable given a cosmic star formation history (CSFH), and what we know about stellar evolution and dust attenuation. Recently, \cite{Driver2012} and \cite{2018MNRAS.474..898A}, built models to predict the EBL (\citealp{2016ApJ...827..108D}), and its subdivision by redshift into CSED slices. These models require as input a CSFH parametrisation, stellar libraries, e.g., \textsc{Pegase2} or BC03 (\citealp{2003MNRAS.344.1000B}), an initial mass function (IMF), and simple axioms around metallicity evolution (typically linked to star-formation), dust attenuation and dust emission, and, in the \cite{2018MNRAS.474..898A} case, the ability to incorporate obscured and unobscured AGN components. Here, we essentially repeat this process, but we now adopt the comprehensive package \textsc{ProSpect} (\citealp{Robotham2020}). This package has been developed to both model and generate individual galaxy SEDs, with significant robustness, flexibility and functionality. It also includes a sub-module to provide an EBL prediction from UV to FIR, given the kind of assumptions described above. \textsc{ProSpect} therefore provides us with the opportunity to use the COB to constrain model inputs to the EBL code, and in particular the CSFH normalisation (see \citealp{Robotham2020}; \citealp{2020arXiv200511917B}). \\ 

The CSFH is a fundamental empirical measurement describing the evolution of the galaxy population from the past to the present time. Empirically the CSFH has been measured via a number of complementary methods i.e. broadband and spectral line measurements (the compilation of measurements by \citealp{2014ARA&A..52..415M}, hereafter MD14); The core sampling technique results from \textsc{MAGPHYS} (\citealp{2018MNRAS.475.2891D}, D18); VHE constraints (\citealp{2018Sci...362.1031F}, FL18); SED forensic reconstruction (\citealp{2020arXiv200511917B}, B20) and IFU forensic reconstruction (\citealp{2019MNRAS.482.1557S}, S19; \citealp{2018A&A...615A..27L}, L18). On the whole the distinct methods for reconstructing the CSFH agree on the overall trend and shape, i.e., a rapid rise to cosmic noon at $z \leq 2$, and thereafter a gentle continuous decline. In detail, the range of CSFH measurements exhibit an uncertainty of almost a factor of 3 at cosmic noon (see Fig. ~\ref{fig:CSFH_before}). In this work we attempt to reduce the uncertainty by including our COB measurements to the modeling as a new constraint.

\subsection{A simple analytical representation of the Cosmic Star-Formation History}
\label{prospect}
To predict the COB using the \textsc{ProSpect} EBL model, we represent each measured CSFH as a simple analytic function for computational purposes. We elect to use the skewed normal function (\textsc{Snorm}), defined in \textsc{ProSpect} as \textbf{massfunc$\_$snorm} function given by Eq \ref{eq-snorm1}, which can be defined by four free parameters:
\begin{equation}
	    \label{eq-snorm1}
	    \begin{aligned}
	        \rm{SFR}(\rm{age})=\texttt{mSFR}\times e^{\frac{-X(\rm{age})^2}{2}}
	    \end{aligned}
\end{equation}

\noindent

where $X$ is age dependent and defined as:

\begin{equation}
	    \label{eq-snorm2}
	    \begin{aligned}
	        X(\rm{age}) = \left(\frac{\rm{age}-\texttt{mPeak}}{\texttt{mPeriod}}\right) {\left(e^{\texttt{mSkew}}\right)}^{\rm{asinh}\left(\frac{\rm{age}-\texttt{mPeak}}{\texttt{mPeriod}}\right)} 
	    \end{aligned}
\end{equation}

\noindent

where ``age'' is in lookback time (Gyrs) and the four free parameters are defined as: 

\begin{itemize}
\item \noindent \textbf{m\textsc{SFR}}:
    Star formation rate normalisation in units of $\rm{M}_{\odot}\rm{yr}^{-1}$; 
\item \noindent \textbf{m\textsc{Peak}}:
    The lookback age of the star formation history peak in Gyr;
\item \noindent \textbf{m\textsc{Period}}:
    Time period of the star formation i.e. the standard deviation of the star formation history period; and 
\item \noindent \textbf{m\textsc{Skew}}:
    The skew of the star formation history (0 for normal, $<$ 0 for long tail to old ages and $>$ 0 for long tail to young ages.)
\end{itemize}

It is noted that m\textsc{SFR} is the peak SFR value and modifying its value scales the entire plot up and down without changing the shape and hence it also acts as a normalisation parameter.\\

\begin{figure*}
    \centering
    \includegraphics[scale = 0.4]{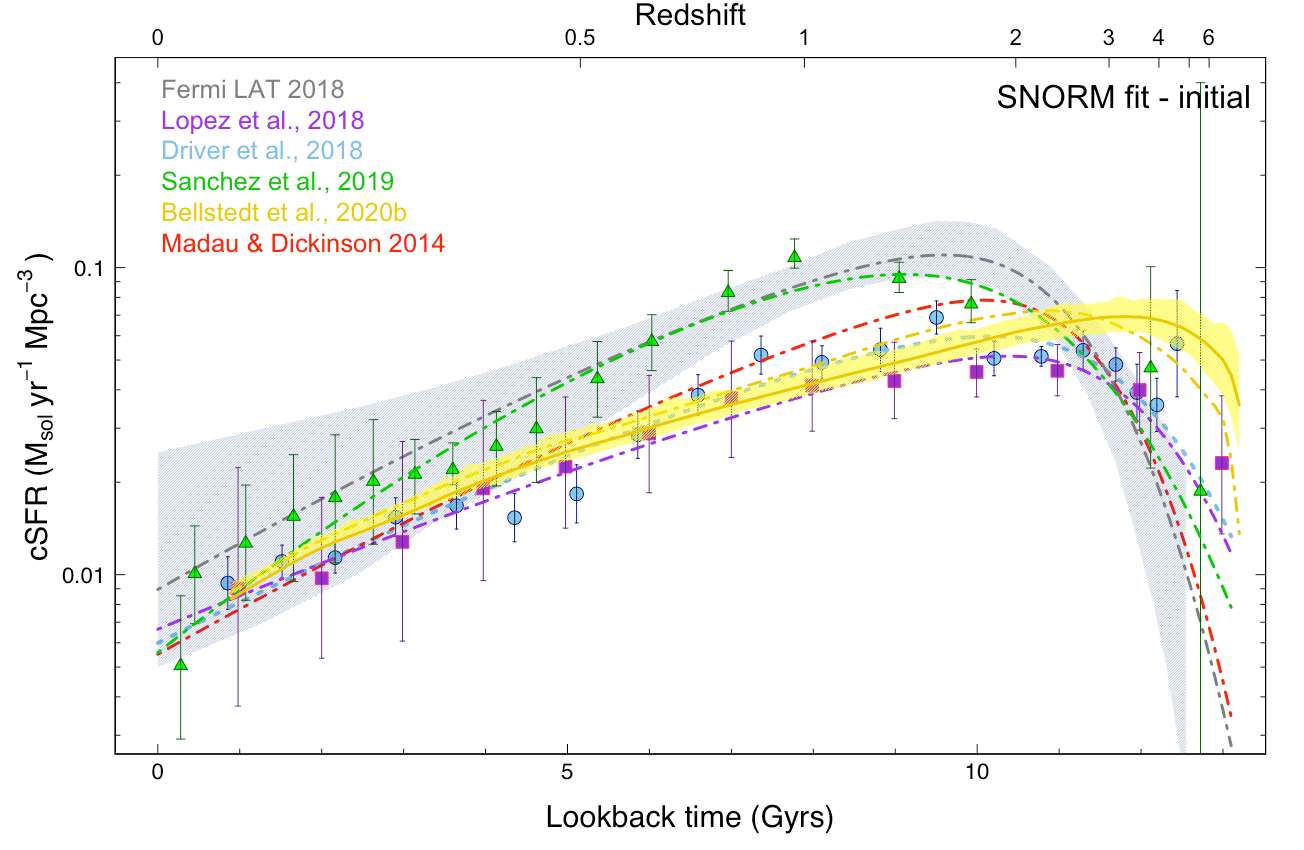}
    \caption{The cosmic star-formation rate before our constraint using mSFR (the peak value) parameter (as described in text) shown for the various methods: SED core sampling in sky-blue (\citealp{2018MNRAS.475.2891D}); VHE constraints in gray (\citealp{2018Sci...362.1031F}); SED forensic reconstruction in gold (\citealp{2020arXiv200511917B}); IFU forensic reconstruction in purple and green (\citealp{2018A&A...615A..27L} and \citealp{2019MNRAS.482.1557S} respectively). We also show the curve from~\protect\cite{2014ARA&A..52..415M} as the dashed red line. The solid gold line shows the raw CSFH value from ~\protect\cite{2020arXiv200511917B}.}
    \label{fig:CSFH_before}
\end{figure*}

We fit each of our adopted CSFHs with the \textsc{Snorm} function and show the fitted values in Table~\ref{table:freeparams} and also plotted as lines in Fig.~\ref{fig:CSFH_before}. The solid gold line shows the raw CSFH values from \cite{2020arXiv200511917B} where the yellow shaded region shows the measurement uncertainty from the sampling of the MCMC chains that determine the SFH. In general the \textsc{Snorm} parametrisation is a reasonable fit to the various data and places all datasets onto a simple comparable footing, with perhaps the exception of the \cite{2019MNRAS.482.1557S} dataset at very low redshift. Note: the \cite{2019MNRAS.482.1557S} study used a Salpeter IMF, and the values have been rescaled to a Chabrier IMF by multiplying by $\times$0.63 (\citealp{2014ARA&A..52..415M}).

%%%%%%%%%%%%%%%%%%%%%%%%%%%%%%%%%%%%%%%
\subsection{EBL predictions with \textsc{ProSpect}}
\label{modelling_EBL}
\begin{figure*}
    \centering
    \includegraphics[width = \textwidth]{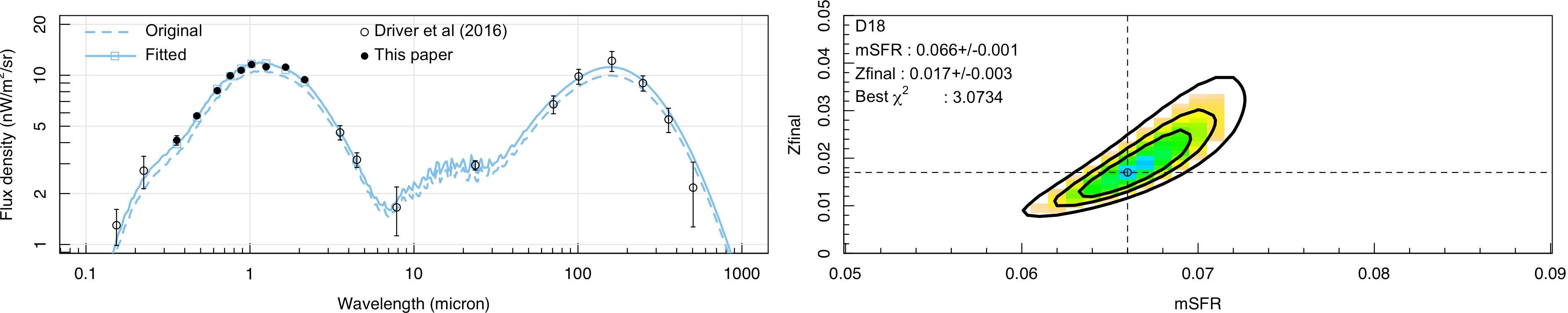}
    \includegraphics[width = \textwidth]{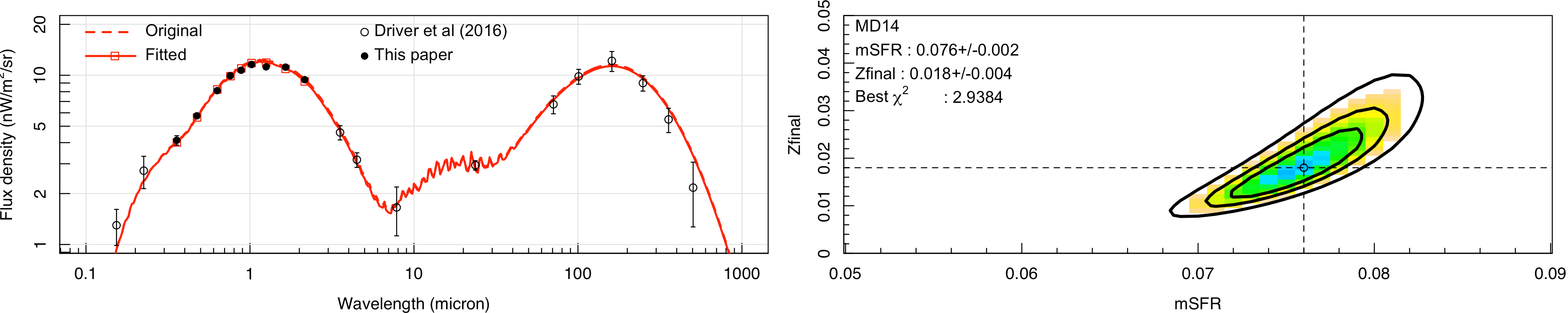}
    \includegraphics[width = \textwidth]{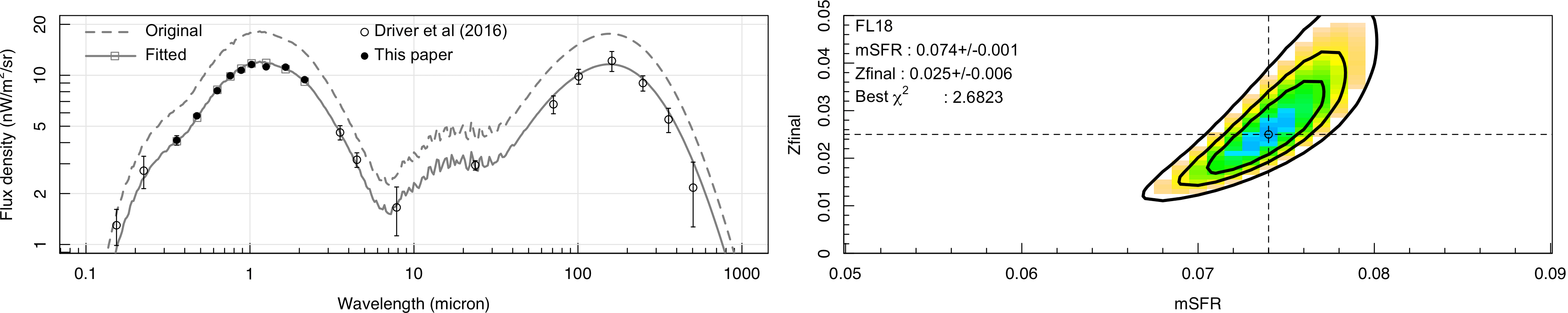}
    \includegraphics[width = \textwidth]{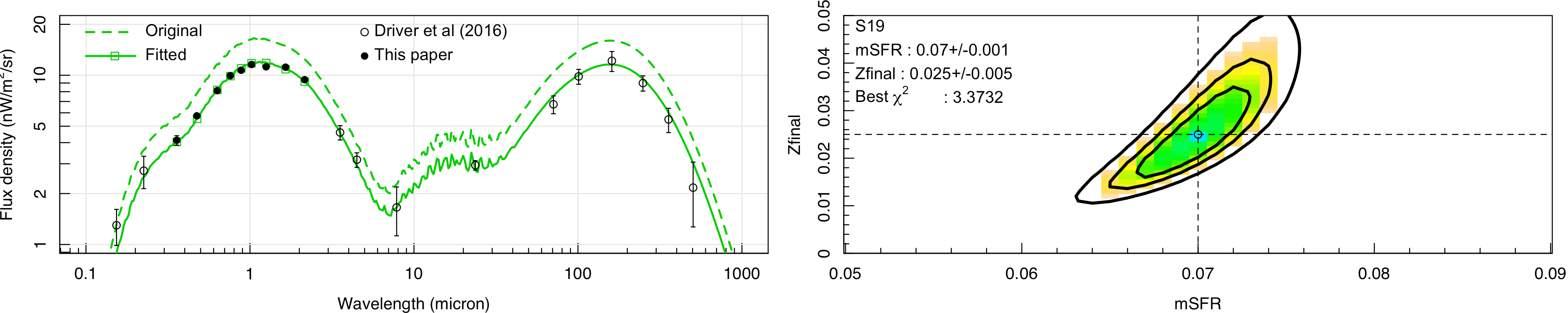}
    \includegraphics[width = \textwidth]{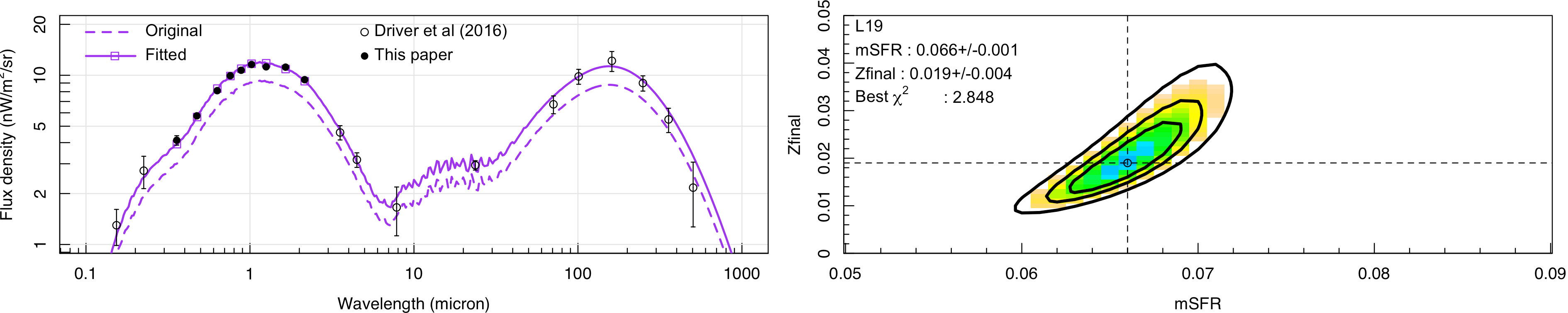}
    \includegraphics[width = \textwidth]{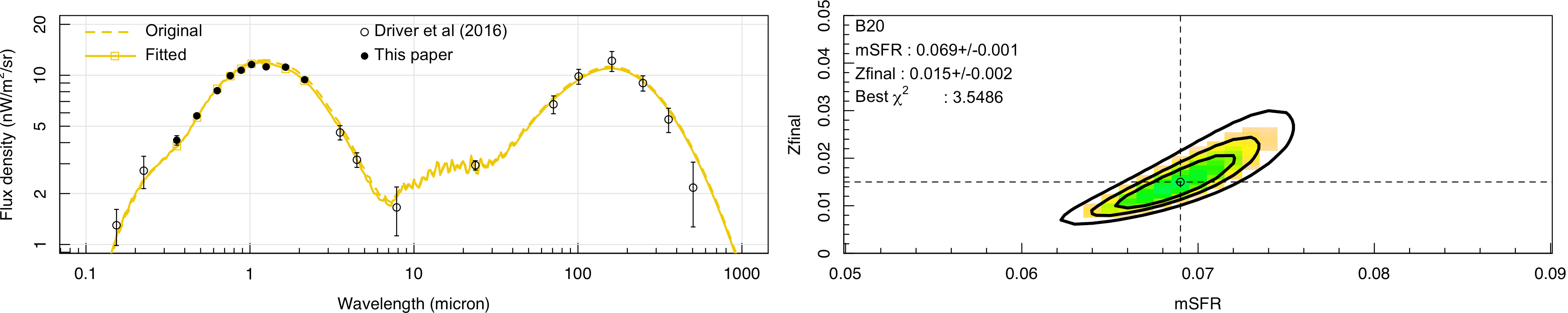}    
    \caption{The reduced $\chi^{2}$ analysis for each of our star-formation histories showing the initial and final EBL model predictions compared to the data, and (right panels) the 1, 2 and 3 $\sigma$ contours in the mSFR-Zfinal plane. The y-axis shows the final metallicities  ($Z_{\rm final}$). This analysis has been summarised in Table \ref{table:freeparams}.}
    \label{fig:Chisq}
\end{figure*}

With a simple representation of the CSFH in place, we can now make our COB predictions using the \textsc{ProSpect} EBL routine (see EBL vignette on github\footnote{https://github.com/asgr/ProSpect}). In brief the \textsc{ProSpect} prediction adopts the Chabrier IMF (\citealp{2003PASP..115..763C}) and defines the COB/CIB by condensing the full galaxy population down to a canonical representative galaxy intended to reflect a mean mass-weighted SED. This canonical representative SED, includes the following elements: 
\begin{itemize}
\item an invariant Chabrier IMF (\citealp{2003PASP..115..763C}).
\item closed box stellar evolution using BC03 (\citealp{2003MNRAS.344.1000B}) simple stellar population libraries.
\item two-components dust model to represent the birth clouds and the inter-stellar medium (ISM)
\item a FUV-IR attenuation prescription following the free form variant of the \cite{2000ApJ...539..718C} prescription (independent for each dust component)
\item energy-conserving flux redistribution to the far-IR using templates by \cite{2014ApJ...784...83D} (independent for each dust component)
\item evolving gas-phase metallicity that tracks star-formation, i.e., grows with the locked-up stellar mass (\citealp{2013MNRAS.430.2622D})

\end{itemize}
In running our model for our canonical galaxy we adopt the opacities ($\tau_{\rm BC}=1$, $\tau_{\rm ISM}=0.3$), temperatures/profile shapes ($\alpha_{\rm BC}=1.75$, $\alpha_{\rm ISM}=3.0$), and the default birth cloud/ISM power laws (-0.7, as recommended by \citealp{2000ApJ...539..718C}). These parameters represent the \textsc{ProSpect} defaults, except for the $\alpha_{\rm BC}$ value which is increased from 1.0 to 1.75 implying slightly cooler dust, and was chosen to shift the predicted FIR dust peak to longer wavelengths to better align with the observed CIB measurements. With these parameters fixed throughout, we can now feed in the parameterised CSFH into the \textsc{ProSpect} EBL routine to produce a full COB/CIB prediction of the extragalactic background light as observed at $z=0.0$. The dashed lines in the left panels of Fig.~\ref{fig:Chisq} show the resulting EBL predictions. Note that, for CSFH measurements that are normalised higher than the others (\citealt{2019MNRAS.482.1557S} and \citealt{2018Sci...362.1031F}), the resulting EBL over-predicts our EBL measurements.

For each of our six CSFH paramaterisations, we now generate a grid of COB/CIB predictions, as viewed through the $ugriZYJHK_s$ filter-set, by fitting for a range of CSFH normalisations (mSFRs), and final metallicities ($Z_{\rm final}$). As discussed in detail by \cite{2020arXiv200511917B}, the SED of a galaxy is influenced by not only the age distribution of the stellar population, but also the metallicity of the stellar population. It is therefore important for us to simultaneously fit for the modelled final gas-phase metallicity ($Z_{\rm final}$), in addition to the normalisation of the CSFH (mSFR). Using the closed-box metallicity evolution prescription within \textsc{ProSpect}, the shape of the metallicity evolution is prescribed by the build-up of stellar mass, but the target final metallicity value is a free parameter. Inherently, this value represents the light-weighted metallicity at which stars are forming at the present day, which could be interpreted as the light-weighted ISM metallicity. We note that this is not a measurement of the overall ``cosmic" gas-phase metallicity, as this approach is only sensitive to star-forming gas. Consequently, the $Z_{\rm final}$ parameter can be regarded as a kind of nuisance parameter, required for fitting in order to accurately model the metallicities of the stellar populations in the galaxy sample.

In order to compute the revised mSFR, we simply scale up and down the mSFR value until we get the best $\chi^{2}$ between the EBL model prediction and the EBL data in the $ugriZYJHK_s$ bands. Each CIB/COB prediction is regressed against the COB data, the error-weighted (reduced) $\chi^2$ values determined for each prediction, and the final best-fit and error contours determined for each of our adopted CSFH representations. The model results, showing the original and final COB/CIB predictions, for each of our adopted CSFHs are shown in Fig.~\ref{fig:Chisq} (left panels), along with the 1$\sigma$, 2$\sigma$ and 3$\sigma$ confidence levels (right panels). 

\begin{figure*}
    \centering
    \includegraphics[scale = 0.4]{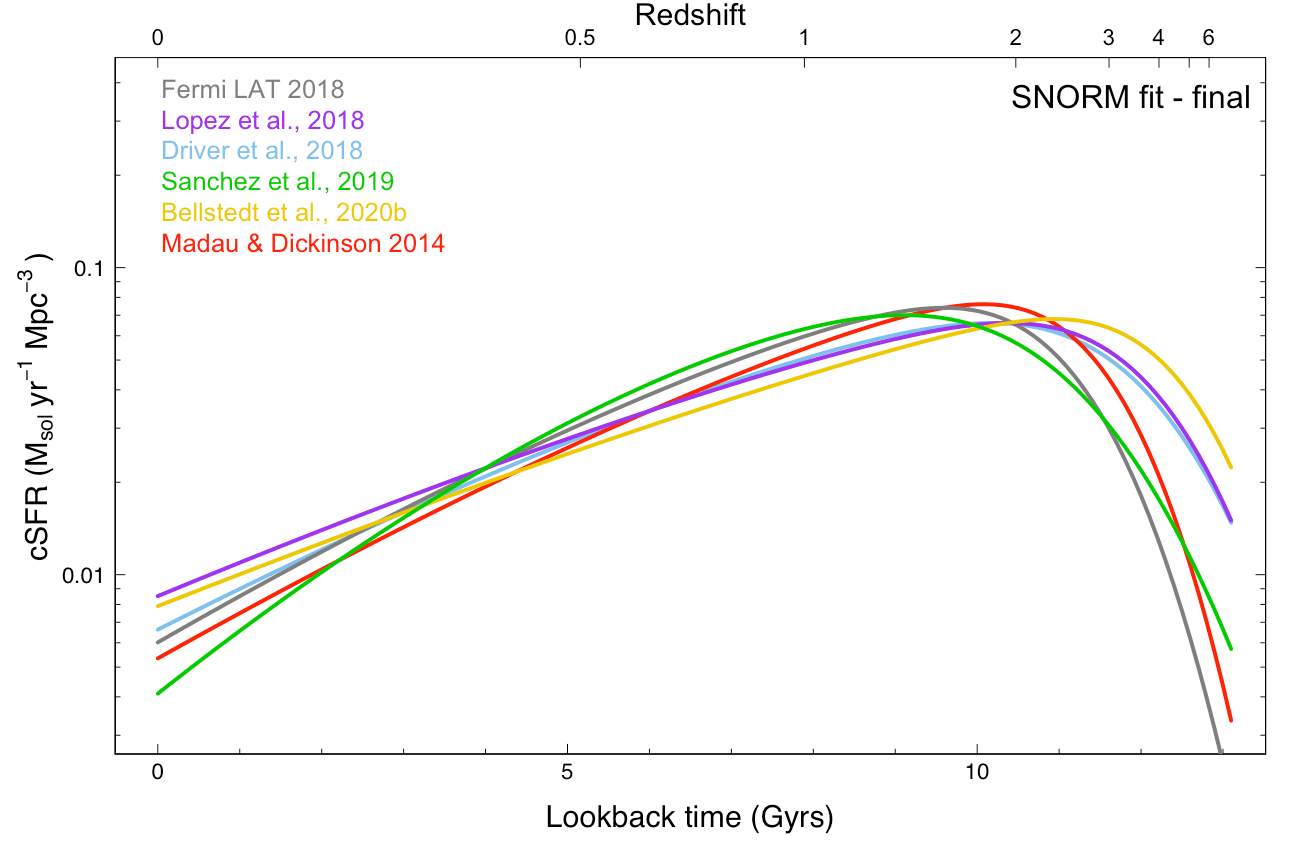}
    \caption{The cosmic star-formation rate after our constraint.}
    \label{fig:CSFH_after}
\end{figure*}

Table~\ref{table:freeparams} summarises the result of the $\chi^{2}$ analysis, showing the four initial free parameters of the \textsc{Snorm} function, the final metallicity, the revised mSFR and best $\chi^2$-value after applying the EBL constraint to each CSFH representation. Note that the reduced-$\chi^2$ values are reasonably distributed around unity, as would be expected if errors are reasonable and the model appropriately describes the data. The revised \textsc{Snorm} values represent a range of adjustments with the CSFHs of MD14 (\citealp{2014ARA&A..52..415M}) and B20 (\citealp{2020arXiv200511917B}) requiring the least re-scaling and essentially fully consistent with their original specified values. The remaining CSFH's all required adjustment to match the COB/CIB data. Figure ~\ref{fig:CSFH_after} shows the set of CSFHs renormalised, and the scatter is notably reduced around cosmic noon from $\times 3$ to $<30$\%.

Following renormalisation against the COB/CIB, we can therefore report a peak value of 0.067-0.075 M$_{\bigodot}$yr$^{-1}$Mpc$^{-3}$ at z $\sim$ 1.5 -- 3. Hence this analysis provides an interesting additional constraint on CSFH measurements and provides strong support for the values reported by MD14 from core-sample style analysis, and the recent forensic measurement from $\sim$7,000 GAMA galaxies by B20. The analysis is in the sense that the MD14 and B20 required the smallest adjustments of their initial mSFR values. It also provides some support for the adopted dust model parameters which well match the far-IR data (see Fig.~\ref{fig:ebl_model}). In future analysis it should be possible to also fit for the dust parameters as our COB/CIB errors improve.

\begin{table*}
    \caption{Free parameters obtained from the \textsc{Snorm} function and our best SFR and metallicity values for six different methods before and after rescaling. \label{table:freeparams}}
    \begin{tabular}{c|c|c|c|c|c|c|c} \hline
   Method & $m\textsc{SFR}_{final}$  & $m\textsc{SFR}_{initial}$ & $m\textsc{Peak}$ & $m\textsc{Period}$ & $m\textsc{Skew}$ & $Z_{best}$ & $chi^2_{best}$ \\ \hline \hline    
   ~\protect\cite{2018MNRAS.475.2891D}; D18 & 0.069$\pm$0.001  & 0.06 & 10.164 & 2.369 & 0.321 & 0.017$\pm$0.003 & 3.07 \\ 
   Madau $\&$ Dickinson (2014); MD14 & 0.076$\pm$0.002  & 0.078 & 10.077 & 1.884 & 0.354 & 0.018$\pm$0.004 & 2.94 \\
   ~\protect\cite{2018Sci...362.1031F}; FL18 & 0.074$\pm$0.001  & 0.110 & 9.594 & 2.010 & 0.334 & 0.025$\pm$0.006 & 2.68 \\ 
   ~\protect\cite{2019MNRAS.482.1557S}; S19 & 0.070$\pm$0.001  & 0.095 & 9.114 & 2.393 & 0.229 & 0.025$\pm$0.005 & 3.37  \\ 
   Lopez Fernandez et al., (2019); L19 & 0.066$\pm$0.001  & 0.051 & 10.378 & 2.286 & 0.364 & 0.019$\pm$0.004 & 2.85  \\
   ~\protect\cite{2020arXiv200511917B}; B20  & 0.069$\pm$0.001  & 0.073 & 10.94 & 2.074 & 0.394 & 0.015$\pm$0.002 & 3.55 \\ \hline
    \end{tabular}
\end{table*}

\begin{table}
    \caption{A sample of our EBL spectrum from \textsc{ProSpect}. The full table in csv format $(file name: EBL\_model.csv)$ is provided electronically as supplementary material. The first column identify the wavelength in angstrom for a range of FUV to FIR and the second column shows the derived flux in the unit of $(nW/m^{2}sr^{-1})$. \label{table:EBLspectra}}
    \centering
    \begin{tabular}{c|c} \hline
   Wavelength (\AA)  & Flux $(nW/m^{2}sr^{-1})$  \\ \hline \hline    
   1122.0185    & 0.2305 \\ 
   1148.1536    & 0.2753 \\ 
   1174.8976    & 0.3160 \\ 
   ...    & ... \\ 
   9549925.9    & 0.6487 \\ 
   9772372.2    & 0.6046 \\ 
   10000000.0   & 0.5636 \\ \hline
    \end{tabular}
\end{table}

\subsection{The stellar mass density}
\begin{figure}
    \centering
    \includegraphics[scale = 0.45]{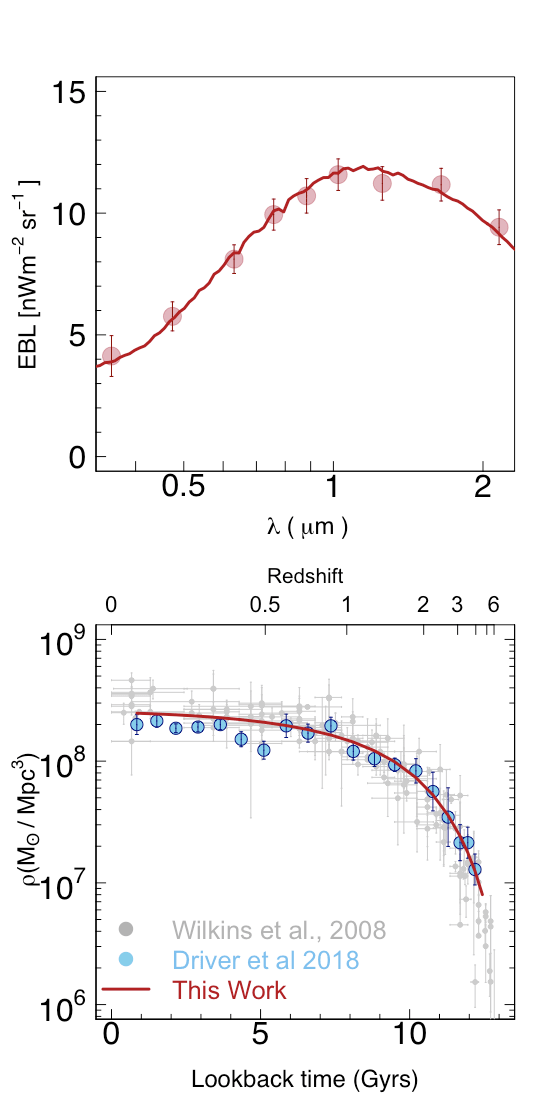}
    \caption{(top) the best fit EBL model from \textsc{ProSpect}; (bottom) the derived stellar mass density compared to data from~\protect\cite{2008MNRAS.385..687W} (gray) and ~\protect\cite{2018MNRAS.475.2891D} (blue).}
    \label{fig:trio}
\end{figure}

The top panel in Fig. \ref{fig:trio} shows a zoom of the best fit EBL from $u$ to $K_s$ band as obtained by \textsc{ProSpect} using the renormalised D18 CSFH. We also show the associated prediction for the stellar mass density (SMD) based on this CSFH (bottom panel).\\

This SMD prediction, shown as the red curve in the bottom panel of Fig.~\ref{fig:trio} compares well to data from \cite{2008MNRAS.385..687W} (grey data points) and D18 (blue data points), where the latter represents the full GAMA/G10-COSMOS/3D-HST combined dataset, spanning almost the entire age of the Cosmos. In detail, we see that the model only just lies within the errorbars of the data which is most likely a reflection of the fact that \textsc{ProSpect} does produce slightly higher stellar masses, given the same data, than MAGPHYS (see \citealp{Robotham2020} figure 33, lower panel) which was used to determine the datapoints in D18 (\citealp{2018MNRAS.475.2891D}). However, there are also two other plausible physical explanations. One is that mass-loss is very slightly higher than assumed for a Chabrier IMF. The second and perhaps more interesting is that such a discrepancy could potentially represent stripped or missing local mass, as the IGL is predominantly derived from light production at intermediate redshifts while the SMD is measured locally (see \citealp{2018PASP..130f4102A} and Ashcraft 2018 PhD dissertation). Hence if some stellar mass is stripped from the progenitors of the present day population in the latter half of the Universe, this could manifest as a modest discrepancy between the EBL predicted SMD and that measured. The obvious way to distinguish, is by breaking the EBL into its constituent CSED time slices, where one might start to identify less energy than expected if mass is being stripped and not accounted for. The reverse is also true, in that a good match between the CSFH, SMD and COB/CIB data suggests, mass stripping is a very small factor (few percent), otherwise the predicted SMD would significantly over-predict the measured stellar-mass density. To bring our predicted SMD perfectly in line with the data, would require and offset of about 20 percent, which is consistent with the 15 per cent stellar mass offset between MAGPHYS and \textsc{ProSpect} (see \citealp{Robotham2020}; \citealp{2020arXiv200511917B}). The implication, is that our results are consistent with mass stripping of anywhere from 0-10 per cent. This constraint is likely to be tightened as we improve both our stellar mass estimates of the nearby population, as well as our COB/CIB constraints.\\

\subsection{The final EBL and comparison to models}
\begin{figure*}
    \centering
    \includegraphics[scale = 0.35]{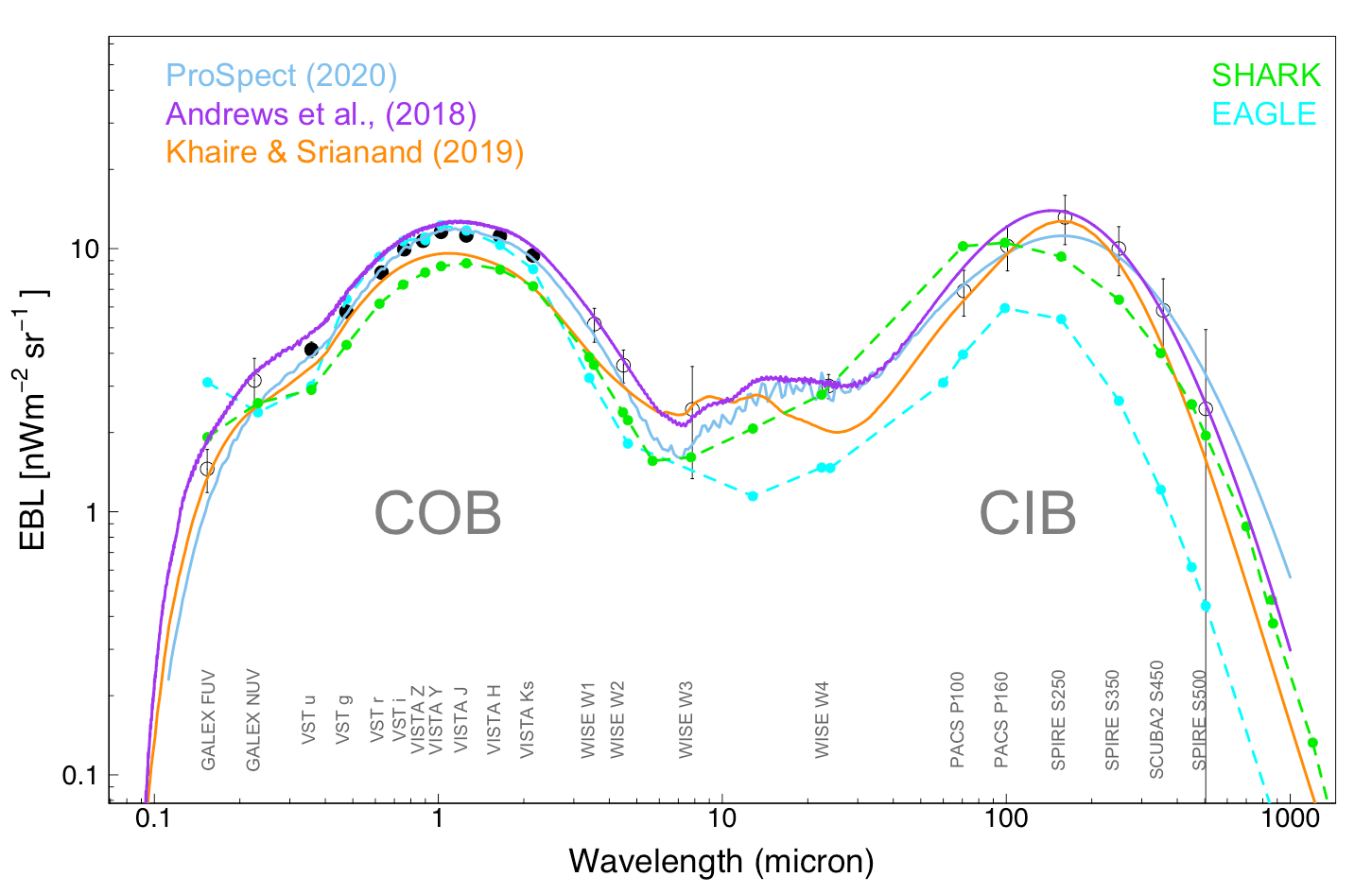}
    \caption{Final EBL model from~\protect\cite{2018MNRAS.475.2891D} in sky blue is compared to the other data from semi-analytical model (green), synthesis model (orange) and phenomenological model (purple). We overplotted the IGL data from $u$ to $K_s$ obtained from our number counts in closed circles combined with the IGL data for UV, mid-IR and far-IR from~\protect\cite{2016ApJ...827..108D} in open circles to present both CIB and COB. EAGLE data is shown in cyan.}
    \label{fig:ebl_model} 
\end{figure*}

Finally, Fig. \ref{fig:ebl_model} presents our best-fit EBL model based on the modified D18 (\citealp{2018MNRAS.475.2891D}) CSFH, to the full COB and CIB data, shown as black data points. Note that the new IGL data from our revised number counts ($ugriZYJHK_s$) are shown as closed black circles, and this is now complemented by our previous IGL results \cite{2016ApJ...827..108D} shown as open black circles to extend the plot to the UV, MIR and FIR regimes. Overlaid are recent model predictions from our earlier phenomenological model (purple line; \citealp{2018MNRAS.474..898A}), data from the new SHARK semi-analytic model (green line; \citealp{2019MNRAS.489.4196L}), and the recent Khaire \& Srianand phenomenological model (orange line; \citealp{2019MNRAS.484.4174K}). All provide reasonable descriptions of the observed distribution, with the optimised \textsc{ProSpect} fit providing the closest match over not just the $u-K_s$ regime but the entire FUV to FIR range. We provide our optimised model in electronic Table form, with Table \ref{table:EBLspectra} showing the first and last three lines of the data file, and advocate the use of these data as the best current representation of the COB and CIB. A related point to consider is that the comparison between SHARK and the observation at the FIR likely requires further analysis.\\

The cyan dashed line shows the EBL calculation from the EAGLE data\footnote{Data obtained from the public database (\citealp{2016A&C....15...72M})}. We compute the predicted EBL from the EAGLE simulations (\citealp{2015MNRAS.446..521S}) by retrieving from the EAGLE public database (\citealp{2016A&C....15...72M}) all galaxies from z=0 to z=9 that had an observer-frame r-band flux $>0$ (i.e. a flux calculation was possible) and their full photometry in the bands GALEX FUV and NUV, SDSS u, g, r, i, z, UKIDSS Z, Y, J, H, K, WISE bands 1, 2, 3 and 4, MIPS 24 microns, IRAS 60 microns, PACS 70, 100 and 160 microns, SPIRE 250, 350 and 500 microns, and SCUBA2 450, and 850 microns. These fluxes were computed via postprocessing of EAGLE galaxies using the radiative transfer code SKIRT (see \citealp{2017MNRAS.470..771T} for details). For each redshift, we compute the implied projected sky area of a $\rm (100\, cMpc)^2$ square assuming the EAGLE cosmology (\citealp{2014A&A...571A..11P}) and the energy contributed by that redshift, $E(z)$, as the sum of the fluxes of all galaxies at that redshift divided by the solid angle corresponding to the area above. We then integrate under the E(z) vs. z curve to obtain the predicted EBL.\\

We note that the fluxes used to compute the predicted EBL for EAGLE, only include galaxies that have a number of (sub-)particles representing the galaxy's body of dust $>250$ (see \citealp{2018ApJS..234...20C} for details). This roughly ignores about $\approx 30$\% of galaxies with stellar masses $>10^{10}\,\rm M_{\odot}$ at $z<0.4$, which are expected to be bright in the optical and NIR given their large stellar mass. We hence consider the presented EBL here as a lower limit for EAGLE, but it's likely that the true EBL in the simulation is only $20-30$\% higher in the optical-NIR, which is still $\approx 50$\% lower than the observations indicate. The MIR-FIR part of the EBL, however, is likely less affected by these limits, and hence the under-prediction here is in tension with observations. Furthermore, there are reasons to believe the simulations are underestimates by 20-30 per cent. However even if one incorporated this correction the simulations would still under-predict the EBL observations by 50 per cent. This is not necessarily surprising given the discrepancy in the predicted number counts of EAGLE and observations in the FIR bands (e.g. \citealp{2019MNRAS.487.3082C}; \citealp{2019A&A...624A..98W}). 

%%%%%%%%%%%%%%%%%%
%%% CONCLUSION %%%
%%%%%%%%%%%%%%%%%%
\section{Conclusions}
\label{section7}
This paper presents a measurement of the optical part of the extra-galactic background light via integrated galaxy number-counts to significantly greater accuracy than previous measurements, enabling us to use the COB data to place constraints onto the CSFH. The main results of this work are summarized as:  

\begin{itemize}
    \item We combined three complementary datasets and derived in relatively wide and deep fields the galaxy counts in the optical and NIR bands ($ugriZYJHK_s$) from the GAMA, DEVILS and various deep ground-based and HST datasets. The compilation of wide and deep datasets allow us to measure the galaxy number counts in a wide magnitude interval (spanning more than 15 magnitudes) across the multi-wavelength filters. Where possible filter conversions have been applied to map our measurements onto the ESO VST/VISTA $ugriZYJHK_s$ system. In general our revised number-counts agree well with other published data in all bands and with significantly improved errors over the critical intermediate magnitude range that dominates the IGL signal. 
    
    \item From the number-count data we determine the contribution in each magnitude interval to the total luminosity density. We fit a smooth spline and extrapolate over the entire magnitude range to derive the integrated flux density, or IGL, of discrete sources from a steradian of sky over the full path-length of the universe and in each filter. Compared to previous results we find a 5-15 per cent increase in our IGL measurements from $u-K_s$. We attribute this increase to several factors which include: \textsc{ProFound}'s propensity to recover closer to total fluxes, appropriate filter transformations, and improved accounting for the area lost to bright stars and their associated halos (ghosting).
    
    \item We compared our optical/NIR IGL measurements to VHE COB (i.e., those from Fermi-LAT, MAGIC, H.E.S.S., and  VERITAS), and direct estimates from space platforms (HST, Pioneer 10/11, New Horizons). Our derived optical/NIR IGL results agree well within the errors with all VHE COB measurements, and those from the deep space based missions (i.e., Pioneer 10/11 and New Horizons). The implication is that significant uncertainty must exist in either or both our understanding of dust in the inner Solar system (zodiacal light), or upper atmosphere Earth-shine in the HST dataset.   
    
    \item The new optical/NIR IGL measurements include an improved error-analysis to manage the uncertainty introduced by cosmic variance. We assessed both systematic (zero-point, filter uncertainty, and cosmic variance), and random (spline fitting and Poisson) errors. In doing so we have reduced the uncertainty from $\sim 20$ per cent to below $\sim 10$ per cent. It can also be inferred that this analysis is now mostly dominated by systematic errors, and in particular cosmic variance at bright and intermediate magnitudes, and positioning issues (i.e., zero-point and filter conversions) at faint magnitudes. With the advent of the Rubin Observatory, Euclid, the Roman Space Telescope and JWST  there is a strong prospect to address these issues and reduce the errors closer to the 1 per cent mark. 
    
    \item To model the EBL, we introduced the \textsc{Snorm} function fitted to the data of \cite{2018MNRAS.475.2891D}. This is a simple four-parameters function that describes the adopted CSFHs extremely well. Using \textsc{ProSpect} we provide a prediction of the EBL and regress against our data to determine the optimal normalisation of the \textsc{Snorm} function and the gas-phase metallicity evolution. The model provides an excellent fit from UV to far-IR and places strict constraints on the normalisation of the cosmic star-formation history with a constraint of $0.066-0.076$ $M_{\bigodot}$yr$^{-1}$ $Mpc^{-3}$ at cosmic noon. The constraint is in close agreement with that provided by Madau \& Dickinson. We conclude that both the revised \textsc{Snorm} function and the \cite{2014ARA&A..52..415M} result provide an excellent description of the CSFH which is fully consistent with both the stellar mass growth and energy production over all time.

\end{itemize}
%%%%%%%%%%%%%%%%%%%%%%%%
%%% ACKNOWLEDGEMENTS %%%
%%%%%%%%%%%%%%%%%%%%%%%%
\section*{Acknowledgements}
    GAMA is a joint European-Australasian project based around a spectroscopic campaign using the Anglo-Australian Telescope. The GAMA input catalogue is based on data taken from the Sloan Digital Sky Survey and the UKIRT Infrared Deep Sky Survey. Complementary imaging of the GAMA regions is being obtained by a number of independent survey programmes including GALEX MIS, VST KiDS, VISTA VIKING, WISE, Herschel-ATLAS, GMRT and ASKAP providing UV to radio coverage. GAMA is funded by the STFC (UK), the ARC (Australia), the AAO, and the participating institutions. The GAMA website is http: //www.gama-survey.org/. Based on observations made with ESO Telescopes at the La Silla Paranal Observatory under program ID 179.A-2004.\\
    DEVILS is an Australian project based around a spectroscopic campaign using the Anglo-Australian Telescope. The DEVILS input catalogue is generated from data taken as part of the ESO VISTA-VIDEO (\citealp{2013MNRAS.428.1281J}) and UltraVISTA (\citealp{2012A&A...544A.156M}) surveys. DEVILS is partly funded via Discovery Programs by the Australian Research Council and the participating institutions. The DEVILS website is https://devilsurvey.org. The DEVILS data are hosted and provided by AAO Data Central\footnote{https://datacentral.org.au/}. Parts of this research were conducted by the Australian Research Council Centre of Excellence for All Sky Astrophysics in 3 Dimensions (ASTRO 3D) through project number CE170100013.\\
    This research is supported by the UPA awarded by the University of Western Australia Scholarships Committee and the awarded by the Astronomical Society of Australia.\\
    This work was supported by resources provided by the Pawsey Supercomputing Centre with funding from the Australian Government and the Government of Western Australia.\\
    MS has been supported by the European Union's Horizon 2020 Research and Innovation programme under the Maria Sklodowska-Curie grant agreement (No. 754510), the Polish National Science Centre (UMO-2016/23/N/ST9/02963), and the Spanish Ministry of Science and Innovation through the Juan de la Cierva-formacion programme (FJC2018-038792-I).

\section{Data availability}
The data underlying this article are derived from three distinct sources as described below:\\
1- The GAMA catalogue (\texttt{GAMAKiDSVIKINGv01.fits}) is publicly available via a collaboration request (http://www.gama-survey.org). \\
2- The DEVILS (http://www.devilsurvey.org) catalogue (\texttt{DEVILSProFoundPhotomv01}) is currently available for internal team use and will be presented in Davies at al in prep, and subsequently publicly releases through data central (https://datacentral.org.au/).\\
3- The HST data is obtained from \cite{2016ApJ...827..108D}.\\
All the data products that are used and/or derived in this work are available on request.

%%%%%%%%%%%%%%%%
% BIBLIOGRAPHY %
%%%%%%%%%%%%%%%%
\bibliographystyle{mnras}
\bibliography{bibliography}

\end{document}